\documentclass[11pt,a4paper]{article}
\usepackage[a4paper, left=2.5cm,right=2.5cm,top=2.5cm,bottom=2.5cm]{geometry}
\usepackage{graphicx} 
\usepackage[T1]{fontenc}
\usepackage{amsmath}
\usepackage{amsfonts}
\usepackage{amssymb}
\usepackage{amsthm}
\usepackage{mathtools}
\usepackage{relsize,scalerel}
\usepackage{graphicx}
\usepackage{fancyhdr}
\usepackage{enumitem}
\usepackage{stmaryrd}
\usepackage[nottoc, notlof, notlot]{tocbibind}
\usepackage[backend=biber,style=alphabetic]{biblatex}
\usepackage[colorlinks=true, allcolors=blue]{hyperref}
\usepackage{dsfont}
\usepackage{enumitem}
\usepackage{euscript}
\usepackage{mathrsfs}
\usepackage{caption}
\usepackage{authblk}
\usepackage{subcaption}
\usepackage{tabularx}
\usepackage{comment}
\usepackage[many]{tcolorbox}    	
\usepackage{multirow}
\usepackage{emptypage}
\usepackage{appendix}
\usepackage{makeidx} 
\makeindex
\usepackage{minitoc} 
\usepackage{soul}
\usepackage{xcolor}
\definecolor{unbleu}{rgb}{0.03, 0.15, 0.4}
\definecolor{monvert}{rgb}{0.0,.5,0.0}
\definecolor{britishracinggreen}{rgb}{0.0, 0.26, 0.15}
\definecolor{monbleu}{rgb}{0,.2,.8}
\definecolor{monautrebleu}{rgb}{0,0.4,.75}
\definecolor{applegreen}{rgb}{0.55, 0.71, 0.0}
\definecolor{monrouge}{rgb}{0.8, 0.0, 0.0} 
\definecolor{cadmiumgreen}{rgb}{0.0, 0.42, 0.24}
\definecolor{royalblue(traditional)}{rgb}{0.0, 0.14, 0.4}
\definecolor{black}{rgb}{0.0, 0.0, 0.0}
\definecolor{sepia}{rgb}{0.44, 0.26, 0.08}
\definecolor{teagreen}{rgb}{0.82, 0.94, 0.75}
\definecolor{yellow-green}{rgb}{0.6, 0.8, 0.2}
\definecolor{azure(colorwheel)}{rgb}{0.0, 0.5, 1.0}
\definecolor{awesome}{rgb}{1.0, 0.13, 0.32}
\definecolor{cadmiumyellow}{rgb}{1.0, 0.96, 0.0}
\definecolor{carrotorange}{rgb}{0.93, 0.57, 0.13}
\definecolor{green-yellow}{rgb}{0.68, 1.0, 0.18}
\definecolor{huntergreen}{rgb}{0.21, 0.37, 0.23}
\definecolor{darkorange}{rgb}{0.85, 0.4, 0.0}
\definecolor{carrotorange}{rgb}{0.93, 0.57, 0.13}
\definecolor{turquoise}{rgb}{0.25, 0.88, 0.82}

\hypersetup{linkcolor=royalblue(traditional),citecolor=cadmiumgreen, urlcolor=azure(colorwheel), colorlinks=true}

\usepackage{pgfplots}
\pgfplotsset{compat=1.18}
\allowdisplaybreaks
\numberwithin{equation}{section}
\input cyracc.def

\makeatletter
\newcommand{\uset}[3][0ex]{%
  \mathrel{\mathop{#2}\limits_{
    \vbox to#1{\kern-6\ex@
    \hbox{$\scriptstyle#3$}\vss}}}}
\makeatother

\definecolor{main}{HTML}{5989cf}    
\definecolor{sub}{HTML}{cde4ff}     

\newtcolorbox{boxH}{
    colback = sub, 
    colframe = main, 
    boxrule = 0pt, 
    leftrule = 6pt 
}

\title{An analytical framework to unify ecological and engineering resilience near critical transitions}

\author[1,2]{Tristan Gamot}
\author[1]{Tom J. M. Van Dooren}
\affil[1]{\small Institut d’Ecologie et des Sciences de l’Environnement de Paris (iEES-Paris), Sorbonne Université, CNRS, INRAe, IRD, Université Paris Créteil, Université Paris cité, 75005 Paris, France}
\affil[2]{Centre de Physique Théorique (CPHT), CNRS, École polytechnique, Institut Polytechnique de Paris, 91120 Palaiseau, France}
\date{}

\begin{document}

\maketitle

\begin{abstract}
The capacity of dynamical systems to resist and recover from perturbations, broadly referred to as resilience, is commonly expressed by two complementary quantities: ecological and engineering resilience. As many complex systems exhibit critical transitions, or tipping points, understanding how these resiliences jointly change nearby them is central to characterising and anticipating such shifts. Here, we develop a theoretical framework that clarifies this for bifurcation-induced tipping, i.e., critical transitions triggered by the crossing of a local bifurcation. Using normal form theory, we derive explicit scaling laws for commonly used resilience metrics as functions of the distance to the bifurcation point in parameter space, and show that these extend to general models up to a scaling factor. They are particularly relevant for detecting tipping, where the relative behaviour of metrics matters more than their absolute values. The rates at which metrics decrease as the bifurcation is approached depend on both the type of bifurcation and the metric considered. Furthermore, our results show that, sufficiently close to a local bifurcation, resiliences are intrinsically linked. Our predictions, which replace previously proposed scalings based on heuristic arguments, are validated for three representative models covering all commonly encountered local bifurcations in one-dimensional systems.
    
\end{abstract}


\newpage
\section{Introduction}
Ecosystems are continually exposed to perturbations, and their capacity to absorb, resist, or recover from these has been formalized through various, sometimes competing, notions of stability and resilience. Since the seminal work of Holling \cite{holling1973resilience}, much effort has been devoted to defining resilience metrics and finding ways to measure them in real systems. Nowadays, two definitions of resilience dominate in the literature: engineering resilience, which is the rate at which a system returns to a reference state after a perturbation \cite{pimm1984complexity}, and ecological resilience, which is the magnitude of disturbance that can be absorbed without switching the system to an alternative state \cite{holling1973resilience}. The former is often seen as a measure of local stability whereas the latter represents non-local stability \cite{holling1996engineering, dakos2022ecological}.

When a dynamical system is modelled by an ordinary differential equation (ODE) and one is interested in studying resilience, a typical assumption is that the ODE admits a potential function, often referred to in the ecological literature as the \textit{stability landscape} \cite{scheffer2009early, scheffer2015generic, vasilakopoulos2015resilience, meyer2016mathematical, van2021unifying, dakos2022ecological, krakovska2024resilience}. This assumption does not hold in general, as most systems with state spaces of dimension two or more cannot be described as gradient systems \cite{nolting2016balls, strogatz2018nonlinear}. Fortunately, there is a natural and well-developed mathematical framework which allows resilience notions based on potentials to be extended in systems of higher dimension: the Freidlin-Wentzell quasi-potential \cite{nolting2016balls}. Yet, in the results presented here, we will restrict ourselves to the one-dimensional case and therefore make use of potential functions.

When a potential function exists, resilience metrics are readily interpretable as properties of the stability landscape. Engineering resilience is intrinsically linked to the immediate neighbourhood of a stable equilibrium and is determined by the local curvature of the potential at that point, quantified by the dominant eigenvalue of the Jacobian matrix. Several indirect metrics of engineering resilience are derived from this local curvature.
By contrast, ecological resilience concerns a broader neighbourhood of the stable equilibrium and is associated with more global geometric properties of the stability landscape. Common metrics include the height of the potential barrier separating basins of attraction and the minimal distance from the stable equilibrium to the boundary of its basin of attraction (hereafter referred to as basin width). A comprehensive review of these metrics for ecological applications is provided in \cite{dakos2022ecological}, and another more extensive review generalizes these concepts to make them applicable to any type of attractor \cite{krakovska2024resilience}.

\begin{figure}[h!]
\centering
  \includegraphics[width=.8\linewidth]{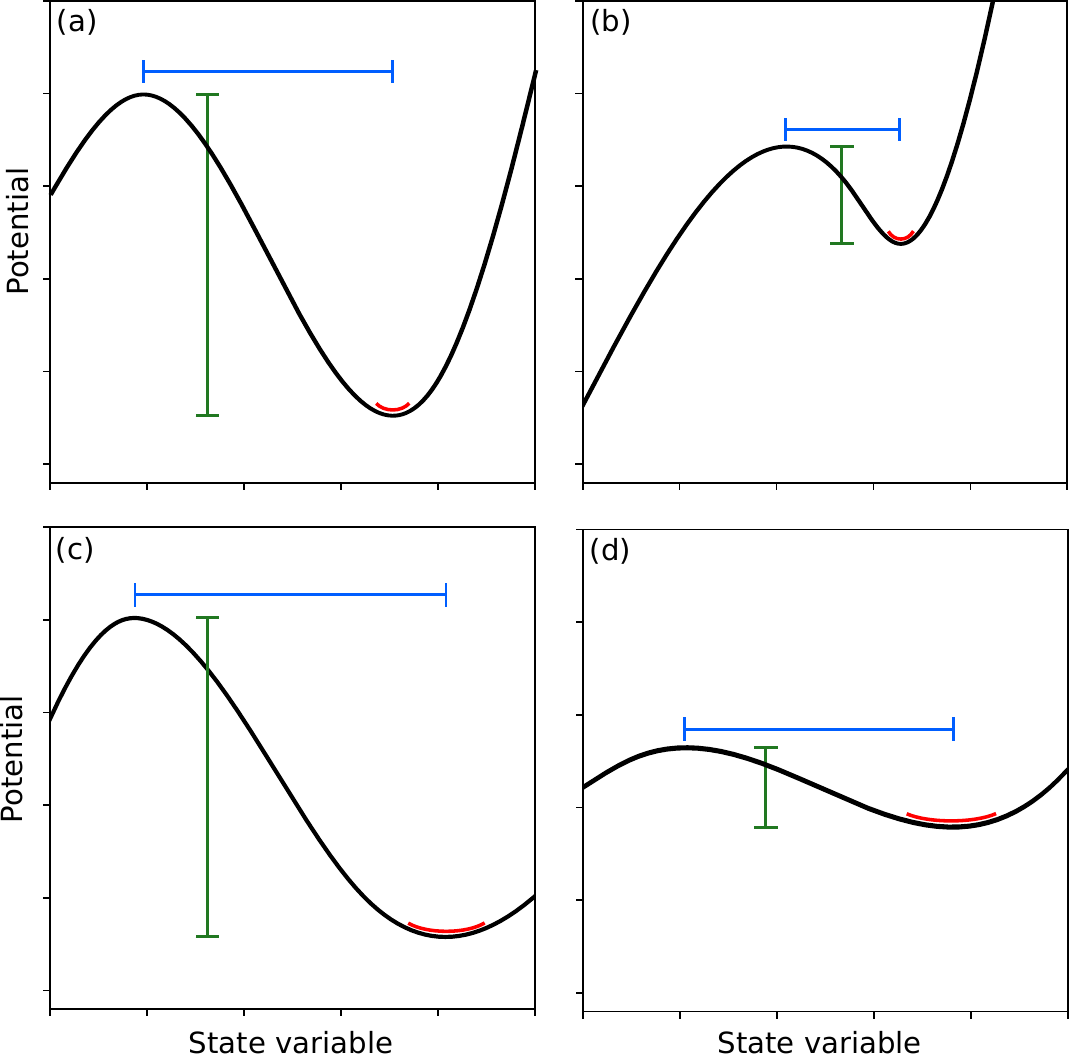}
\caption{Four types of basin of attraction of a potential function around a stable equilibrium (situated at the minimum of the potential), illustrating contrasting combinations of engineering and ecological resilience. In each panel, the dynamical system possesses two equilibria---one stable and one unstable. Red: curvature at the stable equilibrium, used as a measure of engineering resilience. Blue and green: two complementary measures of ecological resilience---basin width and potential barrier height, respectively. (a) High engineering and ecological resilience. (b) High engineering resilience but low ecological resilience. (c) Low engineering resilience but high ecological resilience. (d) Ambiguous ecological resilience: high when measured by basin width, but low when measured by the potential barrier height.}
\label{fig:basin_types}
\end{figure}

Engineering and ecological resilience have long been viewed as fundamentally distinct \cite{holling1996engineering, van2021unifying, dakos2022ecological}. However, one recent study showed that, for various low-dimensional ecological models, the two notions can be strongly correlated \cite{dakos2022ecological}. To support this claim, correlations between ecological and engineering resilience metrics were estimated for random combinations of parameter values, using the potentials of three classical ecological models, a fourth-degree polynomial model, and an exponential model. The authors concluded that these results suggested that engineering resilience can be used as a proxy for ecological resilience, albeit only in specific and relatively simple settings. The case depicted in Figure~\ref{fig:basin_types}a is an example.
They proposed that further research should investigate the conditions under which the found correlations break down.

This naturally raises the question whether scenarios can be envisaged in which engineering and ecological resilience become decoupled. Some possibilities seem obvious. For instance, one may observe a strong curvature of the potential near a stable equilibrium, corresponding to high engineering resilience, together with a low barrier of potential and small basin width, indicating low ecological resilience (Figure~\ref{fig:basin_types}b). Conversely, a low curvature may coexist with a high barrier and large basin (Figure~\ref{fig:basin_types}c). One may even ask whether different metrics of the same notion of resilience can vary independently. For example, a stable equilibrium may be surrounded by a large basin, suggesting high ecological resilience, while the potential barrier is small, indicating low ecological resilience to successive perturbations (Figure~\ref{fig:basin_types}d).

A better characterisation of the different resilience concepts and their interrelation may provide valuable insights across several areas of ecology and evolutionary biology. It could improve the monitoring of losses or gains of resilience under changing conditions in real systems. Resilience monitoring lies at the core of a broad research field which focuses on detecting changes in system dynamics statistics serving as proxies for the distance to a pending critical transition, often termed a tipping point \cite{dakos2008slowing, scheffer2009early}. These indicators, known as early warning signals (EWS) of critical transitions, aim to anticipate abrupt regime shifts, which are often associated with local bifurcations in the deterministic component of a model---a mechanism commonly referred to as bifurcation-induced tipping \cite{ashwin2012tipping}. Within this framework, it is not the absolute magnitude of resilience metrics which is informative, but rather their trends over time, which serve as proxies for the change in resilience as the system approaches or moves away from a critical transition \cite{van2007slow, scheffer2009early}.

According to the foundational literature \cite{van2007slow, dakos2008slowing, scheffer2009early}, the core mechanism underlying EWS is an increasingly slow return to a stable equilibrium following a perturbation when a bifurcation is approached---referred to as critical slowing down---which corresponds to a loss of engineering resilience. Because bifurcation-induced tipping involves local bifurcations, resilience in this context is inherently local: both engineering and ecological resilience primarily characterise the dynamics in a neighbourhood of a stable equilibrium rather than global properties of the phase space.
In this setting, one study \cite{van2007slow} observed already early on that the recovery rate from small perturbations---that is, engineering resilience---can be a remarkably good indicator of ecological resilience, suggesting a strong link between these two classes of metrics. While the ultimate objective of studies on tipping is to detect a loss of ecological resilience, reflected for instance in an increasing vulnerability to shifts towards alternative stable states, such changes are often difficult to measure directly. By contrast, losses in engineering resilience are typically easier to quantify. When the two notions are strongly correlated, monitoring engineering resilience therefore provides an indirect yet practical way to assess ecological resilience. Subsequent studies have implicitly linked engineering resilience to the size of the basin of attraction of the stable equilibrium---an ecological resilience metric---in the context of bifurcation-induced tipping (see Figure~1 of \cite{scheffer2009early} and Figure~1 of \cite{dakos2010spatial}), without investigating under which conditions and why such a link would be expected.

A first attempt to clarify the relationships between engineering and ecological resilience metrics under bifurcation-induced tipping was made in Box~1 of \cite{dakos2015resilience}, where relationships between several metrics, as well as scaling laws describing their dependence on the distance to the bifurcation point, were proposed. However, these results rested on heuristic arguments and the relationships were not explored rigorously.
More recently, \cite{dakos2022ecological} distinguished between resilience metrics based on perturbations of state variables---which leave the potential unchanged---and those based on parameter perturbations, which modify the potential through changes in external conditions and may lead to a bifurcation. Previous research had already considered situations combining both types of perturbations, but focused exclusively on their consequences for engineering resilience metrics, namely the variance and autocorrelation in time series of state variables \cite{dakos2012robustness}.

The analysis of Dakos and Kéfi \cite{dakos2022ecological} was restricted to bistable models which did not necessarily undergo a bifurcation. Here, we instead clarify the links between engineering and ecological resilience metrics associated with state-variable perturbations, as a function of the distance to a bifurcation point in parameter space.
Whereas perturbations of state variables and perturbations in parameter space are treated separately in \cite{dakos2022ecological}, we integrate both within a unified framework, namely that used to study EWS of bifurcation-induced tipping. In this respect, the results presented here provide a theoretical extension of ideas initially introduced in \cite{dakos2015resilience}. We argue that within the framework of bifurcation-induced tipping, the relationship between ecological and engineering resilience can be made explicit. In particular, we show that near local bifurcations, both notions should be viewed as complementary descriptions of a shared local structure determining the dynamics.

To this end, we use the normal form theory of codimension-one local bifurcations in continuous dynamical systems, a framework widely used in the EWS literature. Within this framework, we derive the scalings of commonly used ecological and engineering resilience metrics with respect to the distance to the bifurcation point in parameter space (section~\ref{sec:links_eco_eng_resilience}). This analysis yields explicit relationships between different resilience metrics for each type of bifurcation. We then explain why these power-law scalings extend to more general models which can be reduced to their normal forms (section~\ref{sec:normal_form}). Using three illustrative examples, we show that the theoretical predictions are consistent with both numerical and analytical results (section~\ref{sec:examples_validation}). We then compare the scaling laws obtained here with those reported in previous studies, highlighting agreements and discrepancies (section~\ref{sec:comparison_scalings}). Finally, we discuss the implications of our findings for the measurement and interpretation of resilience in systems subject to critical transitions (section~\ref{sec:discussion}).

\section{Engineering and ecological resilience metrics for normal forms}\label{sec:links_eco_eng_resilience}
\subsection{Use of normal forms}
The analytical study of nonlinear ODEs is often challenging. Nevertheless, an appropriate transformation of coordinates can substantially facilitate their analysis. Introduced by Henri Poincaré \cite{gaeta2002poincare}, normal form theory provides a systematic framework for such transformations. It guarantees the existence of a change of variables---typically restricted to polynomial transformations---that simplifies the functional expressions of the system while preserving the essential features of the flow.

For a parametrized system of the form $\frac{\mathrm{d}x}{\mathrm{d}t} = f(x,r)$, with $f$ smooth in both $x$ and $r$, conditions on the partial derivatives of $f$ can be derived, which hold true when the system undergoes a given type of local bifurcation \cite{wiggins2003introduction}. Normal form theory exploits this structure to reduce local bifurcations to a small set of canonical forms. These simplified representations retain the topological properties of the original system, ensuring that the qualitative dynamics are unchanged, while making the local behaviour more tractable. As these derivatives constrain the leading-order nonlinearities near the bifurcation, they also shape the local geometry of the flow, suggesting that resilience metrics should exhibit bifurcation-dependent behaviour.

In practice, however, the transformation to normal form is carried out only up to a finite order, resulting in an approximate normal form with a controllably small remainder \cite{wiggins2003introduction, strogatz2018nonlinear}. Near an equilibrium, this naturally leads to the systematic elimination of non-essential terms in the Taylor expansion of the scalar field up to a prescribed order, thereby removing higher-order nonlinearities that have only a marginal influence on the local dynamics. 

The use of normal forms to derive generic models for several types of local bifurcations commonly investigated in EWS studies is already well established in the EWS literature \cite{kuehn2011mathematical, boettiger2012quantifying, o2018stochasticity, bury2020detecting, bury2021deep, dylewsky2024early, donovan2024characterising}. 
Since EWS are fundamentally aiming to detect losses of resilience, this framework provides a natural and rigorous setting to investigate the relationship between ecological and engineering resilience. To the best of our knowledge, this connection has not been explicitly explored within a normal form framework.

A fundamental tool for the bifurcation analysis of generic $n$-dimensional ODEs is the centre manifold reduction \cite{crawford1991introduction, kuznetsov1998elements}, which allows to reduce the original system to a lower-dimensional ODE, on the centre manifold, without altering its local bifurcation structure. Regarding the link with normal forms, the common practice is to compute the centre manifold reduction first, and then to compute the normal form for this reduced system (see \cite{glendinning2022normal}, chapter 3 of \cite{guckenheimer1983nonlinear}, or chapter 5 of \cite{murdock2003normal}). However, since many ecological models are formulated as simplified one-dimensional systems and the examples investigated here are such models, this reduction is not required here. We therefore focus on the three most commonly studied one-dimensional bifurcations in continuous-time: the fold (or saddle-node), pitchfork, and transcritical bifurcations \cite{wiggins2003introduction}.

\subsection{Resilience metrics for normal forms}

We are interested in finding the explicit dependencies of the resilience metrics on the bifurcation parameter. These describe how resiliences change as a bifurcation is approached, and contribute to the associated patterns in EWS. For example, starting with the fold normal form, that is 
\begin{equation}
    \frac{\mathrm{d}x}{\mathrm{d}t}=f(x,r)= -r-x^2=- \frac{\mathrm d V_r }{\mathrm d x} (x),
\end{equation}
where $V_r(x)=rx+x^3/3$ is the potential (or \textit{stability landscape}) of the fold normal form at a fixed $r$-value. The fold bifurcation occurs at $r=0$ and for a given $r$-value, with $r<0$, there exist two equilibria, $x^*_{\text{stab}} =+\sqrt{-r}$ (locally stable) and $x^*_{\text{unstab}} =-\sqrt{-r}$ (unstable).

One of the most often used metrics of engineering resilience is the asymptotic exponential rate at which a system returns to its stable state following a small perturbation. In the one-dimensional case, this rate is determined by the local curvature of the potential at the equilibrium. It equals  the negative of the (unique, real-valued) eigenvalue $\lambda$ of the stable equilibrium, with
\begin{equation}\label{eq:dom_ev}
    \lambda= \partial_x f(x^*_{\text{stab}},r)= - \frac{\mathrm d^2 V_r }{\mathrm d x^2}(x^*_{\text{stab}}).    
\end{equation}
Note the sign difference in comparison to Equation~(4) of \cite{dakos2022ecological}. Therefore, for the fold normal form we find that the eigenvalue of $x^*_{\mathrm{stab}}$ is:
\begin{equation}\label{eq:vp_dominante}
 \lambda = -2 \sqrt{-r}= -2\sqrt{\lvert r \rvert} = O(\lvert r \rvert^\alpha).
\end{equation}
where $\alpha =1/2$ is called the recovery exponent \cite{kuehn2011mathematical}. Note that $\lambda<0$ as $x^*_{\text{stab}}$ is stable. The speed of convergence to the stable state is given by the amplitude of $\lambda$, that is $-\lambda$. An equivalent resilience indicator is the characteristic return time \cite{krakovska2024resilience}, which is $-1/\lambda$ in the case considered here. In systems of dimension two or higher that have reactive equilibria---equilibria at which perturbations may transiently amplify before eventually decaying, a property quantified by the reactivity \cite{neubert1997alternatives}---these two resilience measures omit such transient changes following a perturbation and have been criticized for this reason.

As the eigenvalue $\lambda$ quantifies the local curvature of the potential in the neighbourhood of the stable equilibrium (see Appendix \ref{an:curvature} for more details), it also governs the statistical properties of the system in the presence of stochastic perturbations. Such stochastic perturbations allow for an indirect estimation of this curvature. In particular, when noise is added to the dynamics, the dominant eigenvalue controls the stationary distribution of the process, whenever such a distribution exists. Consider for instance the classical case of additive Gaussian white noise,
\begin{equation}
	\mathrm{d}x=f(x,r) \mathrm{d}t + \sigma \mathrm{d}W(t), \label{eq:variance_OU}
\end{equation}
where $\sigma>0$ denotes the noise intensity and $W(t)\sim \mathcal N(0,t)$ is a standard Brownian motion. Linearizing the system around the stable equilibrium $x^*_{\text{stab}}$ leads to an Ornstein–Uhlenbeck process. Its stationary distribution can be derived analytically \cite{gardiner2009stochastic}. In particular, the stationary distribution is Gaussian, with variance given by
\begin{equation}\label{eq:var}
	\mathrm{Var}(x) = \frac{-\sigma^2}{2 \lambda},
\end{equation}
and the autocorrelation function at lag-$\tau \ge 0$ reads
\begin{equation}\label{eq:ac}
	\mathrm{Corr}_{\tau}(x) = e^{\lambda \tau}.
\end{equation}
Note the factor two difference in the exponent in Equation~\eqref{eq:variance_OU} in comparison to Equation~(7) of \cite{dakos2022ecological}.

Under these stochastic assumptions, the entire autocorrelation structure of the process is therefore determined by the single quantity $\lambda$. As a consequence, the autocorrelation at lag-1 is commonly used as a practical proxy for engineering resilience, as it is both theoretically justified and relatively easy to estimate from empirical time series. Although the two equations \eqref{eq:var} and \eqref{eq:ac} (at lag-1 for the autocorrelation) appear in \cite{dakos2022ecological}, we have recalled here the assumptions required for their derivation and interpretation as engineering resilience metrics.

Regarding ecological resilience measures, the potential depth $h$ and basin width $ds$ are widely used \cite{dakos2022ecological}. The former, namely the height of the potential barrier, can be used to estimate the average time the process remains in the basin of attraction of the stable equilibrium before escaping for the first time \cite{gardiner2009stochastic}. The latter, the minimum distance to the basin boundary, measures the maximal amplitude of a single perturbation that the system can handle without leaving the basin of attraction of the stable state. In the fold normal form model, we find that 
\begin{align}
    h &=V_r(x^*_{\text{unstab}}) -  V_r(x^*_{\text{stab}}) \label{eq:barrier_pot} \\
    &= r(-\sqrt{-r})+\frac{(-\sqrt{-r})^3}{3} - \left( r\sqrt{-r}+\frac{(\sqrt{-r})^3}{3} \right) \\
    &= \frac{4}{3}(-r)^{3/2}.    
\end{align}

Regarding the basin width $ds$, we find that 
\begin{align}
    ds &=\lvert x^*_{\text{unstab}} -  x^*_{\text{stab}} \rvert \label{eq:width_basin} \\
    &=2\sqrt{-r}.
\end{align}

Similar calculations for the transcritical and (supercritical or subcritical) pitchfork bifurcations lead to the overview of metrics in Table \ref{tab:resilience_metrics}. We note that the power-law dependence on the distance to the bifurcation point in parameter space ($\lvert r \rvert $ in the normal forms) differs across bifurcation types, which is consistent with the fact that each bifurcation is governed by distinct nonlinear terms. We note that only combinations of ecological and engineering resilience metrics permit distinguishing bifurcation types based on the behaviour of metrics when approaching a bifurcation point.

\begin{table}
\centering
\begin{tabular}{c|c|c|c|c}
  & Resilience metrics & Fold  & Transcritical & Pitchfork \\ \hline
\multirow{2}{*}{Engineering res.} & $\lvert \lambda \rvert$ &   $2\lvert r \rvert^{1/2}$   & $\lvert r \rvert$ & $\lvert r \rvert$ (sub) or $2\lvert r \rvert$ (super)  \\
 & $\mathrm{Corr}_{1}(x)$& $e^{-2\lvert r \rvert^{1/2}}$ & $e^{-\lvert r \rvert}$& $e^{-\lvert r \rvert}$ (sub) or $e^{-2\lvert r \rvert} $ (super) \\
 & $\mathrm{Var}(x)$&   $\frac{\sigma^2}{4\lvert r \rvert^{1/2}}$   & $\frac{\sigma^2}{2 \lvert r \rvert}$ & $\frac{\sigma^2}{2 \lvert r \rvert}$ (sub) or $\frac{\sigma^2}{4 \lvert r \rvert}$ (super) \\ \hline
\multirow{2}{*}{Ecological res.} & $h $      &   $\frac{4}{3}\lvert r \rvert^{3/2}$   &    $\frac{1}{6}\lvert r \rvert^{3}$           &     $\frac{1}{4}\lvert r \rvert^{2}$    \\
                  & $ds$     &   $2\lvert r \rvert^{1/2}$   &      $\lvert r \rvert$         &     $\lvert r \rvert^{1/2}$     \\ \hline
\end{tabular}
\caption{Dependences of the main engineering and ecological resilience metrics on the bifurcation parameter $r$, in the case of the normal forms of the fold, transcritical, (supercritical or subcritical) pitchfork local bifurcations. 
$\lambda$ is the eigenvalue of the stable equilibrium, $\mathrm{Corr}_{1}(x)$ and $\mathrm{Var}(x)$ are the stationary lag-1 autocorrelation and variance of the linearized process, $ds$ is the minimum distance from the stable equilibrium to its basin boundary, and $h$ is the height of the potential barrier separating the stable equilibrium from the alternative one.
Note that $|\lambda|$ is used for the sake of clarity as $\lambda$ is always negative for stable states. The bifurcation point is located at $r=0$. Resilience metrics are each time computed within the basin of attraction of the stable equilibrium, also when an unstable equilibrium exists next to the stable one. For example, in the case of a subcritical pitchfork bifurcation, this implies calculations for $r<0$. When $r>0$, the system admits a single equilibrium, which is unstable, and resilience metrics such as $h$ or $ds$ are therefore not defined.}\label{tab:resilience_metrics}
\end{table}

\section{Relating resilience metrics in normal forms and general models}\label{sec:normal_form}
We aim to clarify how resilience metrics derived from normal form models can be used as proxies for resilience metrics in general models, using truncated normal form reductions. In the following, we consider a possibly complex model of the form $\frac{\mathrm{d}\eta}{\mathrm{d}t} =f(\eta,\alpha)$ which exhibits one of the three previously introduced local bifurcations. Note that the notation $(\eta, \alpha)$ is chosen to distinguish these variables from the normal form variables $(x, r)$ used in section~\ref{sec:links_eco_eng_resilience}. We assume that $f$ is sufficiently smooth to admit a Taylor expansion to arbitrary order. The following conditions guarantee that a non-hyperbolic fixed point exists at the bifurcation point: 
\begin{align}
    f(\eta^*,\alpha^*)&=0, \label{eq:non_hyperbolic_equ_1}\\
    \partial_\eta f(\eta^*,\alpha^*)&=0,\label{eq:non_hyperbolic_equ_2} 
\end{align}

\subsection{Fold bifurcation}\label{sec:normal_form_reduction_fold}
We first consider the case where $f$ exhibits a fold bifurcation at $(\eta^*, \alpha^*)$, meaning that on top of \eqref{eq:non_hyperbolic_equ_1} and \eqref{eq:non_hyperbolic_equ_2}, the following conditions are satisfied (see Chapter 3.3 of \cite{kuznetsov1998elements} or Chapter 20.1C of \cite{wiggins2003introduction}):
\begin{align}
    \partial_\alpha f(\eta^*,\alpha^*)&\neq0, \label{eq:non_hyperbolic_equ_3} \\ 
    \partial_{\eta \eta} f(\eta^*,\alpha^*)&\neq 0. \label{eq:non_hyperbolic_equ_4}
\end{align}
These are genericity conditions (nondegeneracy and transversality), ensuring that the bifurcation is a fold bifurcation.

Expanding $f$ in a Taylor series around $(\eta^*, \alpha^*)$, yields the correspondence
\begin{align}
    \frac{\mathrm{d}\eta}{\mathrm{d} t} &= (\alpha-\alpha^*)\partial_\alpha f(\eta^*,\alpha^*) + \frac{(\eta-\eta^*)^2}{2}  \partial_{\eta \eta} f(\eta^*,\alpha^*)\nonumber
    \\ & \quad + O\big((\alpha-\alpha^*)^2,\,(\alpha-\alpha^*)(\eta-\eta^*),\,(\eta-\eta^*)^3\big), \label{eq:fold_DL}
\end{align}
which, for $\alpha-\alpha^*$ and $\eta-\eta^*$ sufficiently small, can be approximated by
\begin{align}
    \frac{\mathrm{d}\eta}{\mathrm{d} t} \approx (\alpha-\alpha^*)\partial_\alpha f(\eta^*,\alpha^*) + \frac{(\eta-\eta^*)^2}{2} \partial_{\eta \eta} f(\eta^*,\alpha^*). \label{eq:fold_approx}
\end{align}

The relevance of this approximation follows from the fact that the arguments of $O$, which are collections of terms of a specific order, contribute only marginally to the local dynamics near the bifurcation point. This can be shown for each argument in $O$ separately. For example, the $(\eta-\eta^*)^3$ terms are always negligible compared to $(\eta-\eta^*)^2$. If the $(\eta-\eta^*)^3$ terms are not negligible relative to $\alpha-\alpha^*$ (since $\alpha-\alpha^*$ and $\eta-\eta^*$ may be of different orders of magnitude), then by transitivity $\alpha-\alpha^*$ is negligible compared to $(\eta-\eta^*)^2$, so the first term in equation~\eqref{eq:fold_DL} becomes negligible compared to the second and thus $(\eta-\eta^*)^3$ terms don't need to be included in the approximation. Similar reasoning applies to the other arguments, so equation~\eqref{eq:fold_approx} captures the essential behaviour of equation~\eqref{eq:fold_DL} for $\alpha-\alpha^*$ and $\eta-\eta^*$ sufficiently small.

If we implement the following change of variable:
\begin{align}
    x &=\frac{-\partial_{\eta \eta} f(\eta^*,\alpha^*)}{2}(\eta-\eta^*) , \label{eq:change_var_1}\\
    r &= \frac{\partial_{\alpha} f(\eta^*,\alpha^*) \partial_{\eta \eta} f(\eta^*,\alpha^*)}{2} (\alpha-\alpha^*), \label{eq:change_var_2}
\end{align}
we obtain the following approximate ODE for $x$:
\begin{equation}
    \frac{\mathrm{d}x}{\mathrm{d} t} \approx -r-x^2, \label{eq:fold_normal_form}
\end{equation}
which is the normal form of the fold bifurcation. One can consult chapter 20.1c of \cite{wiggins2003introduction} for a rigorous proof that the arguments in the $O$ of equation \eqref{eq:fold_DL} can be neglected and that the dynamics would not be qualitatively changed.

Since these transformations (equations~\eqref{eq:change_var_1}--\eqref{eq:change_var_2}) amount to rescaling both the state variable and the bifurcation parameter distances to the bifurcation point by constant factors, the resilience metrics for the model of equation~\eqref{eq:fold_approx} share the functional form of the dependence on the distance to the bifurcation point (i.e. $\lvert \alpha - \alpha^* \rvert$) with the normal form model of equation~\eqref{eq:fold_normal_form} (i.e. $\lvert r \rvert$; listed in the third column of Table~\ref{tab:resilience_metrics}), up to a constant prefactor.

This transformation of both variables is the simplest possible; higher-order Taylor expansions and therefore polynomial transformations would yield correction terms of arbitrarily high order. Such transformations would refine but not modify the leading-order scaling laws derived here: since the transformation remains polynomial, lower-order terms dominate when the bifurcation parameter and state variable are sufficiently small. The leading-order scalings are recovered in this limit. 


\subsection{Pitchfork bifurcation}
We now consider the case where $f$ exhibits a pitchfork bifurcation at $(\eta^*, \alpha^*)$. In a neighbourhood of $(\eta^*, \alpha^*)$, $f$ satisfies conditions \eqref{eq:non_hyperbolic_equ_1}--\eqref{eq:non_hyperbolic_equ_2} and is odd with respect to $\eta$, i.e., 
\begin{equation}
    f(-\eta, \alpha)=-f(\eta, \alpha).
\end{equation}
For the purpose of genericity, we assume that the following derivatives which are not forced to vanish when partial derivatives of increasing order are inspected, are nonzero:
\begin{align}
    \partial_{\eta \alpha} f(\eta^*,\alpha^*)&\neq 0, \\
    \partial_{\eta \eta \eta} f(\eta^*,\alpha^*)&\neq 0.
\end{align}

Thus, using the fact that $f$ is odd with respect to $\eta$ and expanding it in a Taylor series around $(\eta^*, \alpha^*)$, yields:
\begin{align}
    \frac{\mathrm{d}\eta}{\mathrm{d} t} &= (\eta-\eta^*)\Big((\alpha-\alpha^*)\partial_{\alpha \eta} f(\eta^*,\alpha^*) + \frac{(\eta-\eta^*)^2}{6}  \partial_{\eta \eta \eta} f(\eta^*,\alpha^*)  \nonumber
    \\ & \quad + O\big((\alpha-\alpha^*)^2,\,(\alpha-\alpha^*)(\eta-\eta^*)^2,\,(\eta-\eta^*)^4\big)\Big), \label{eq:pitchfork_DL}
\end{align}
which, for $\alpha-\alpha^*$ and $\eta-\eta^*$ sufficiently small, can be approximated by:
\begin{align}
\frac{\mathrm{d}\eta}{\mathrm{d} t} \approx (\eta-\eta^*)(\alpha-\alpha^*)\partial_{\alpha \eta} f(\eta^*,\alpha^*) + \frac{(\eta-\eta^*)^3}{6}  \partial_{\eta \eta \eta} f(\eta^*,\alpha^*). \label{eq:pitchfork_approx}
\end{align}

A similar reasoning as for the fold bifurcation applied to each argument of $O$ leads to the conclusion that equation~\eqref{eq:pitchfork_approx} captures the essential behaviour of equation~\eqref{eq:pitchfork_DL} for $\alpha-\alpha^*$ and $\eta-\eta^*$ sufficiently small. For a rigorous proof, see chapter 20.1E of \cite{wiggins2003introduction}.

Finally, as above, by applying a transformation where $\eta-\eta^*$ and $\alpha-\alpha^*$ are multiplied by constants where new variables are denoted $(x,r)$ yields the normal form of the pitchfork bifurcation:
\begin{equation}
    \frac{\mathrm{d}x}{\mathrm{d} t} \approx r x \pm x^3, \label{eq:pitchfork_normal_form}
\end{equation}
where the sign $\pm$ depends on whether the pitchfork bifurcation is subcritical or supercritical.

Once again, we recover the same dependencies of the resilience metrics on the distance to the bifurcation point in both the general model and the normal form, at least sufficiently close to the bifurcation.

\subsection{Transcritical bifurcation}\label{sec:normal_form_reduction_transcritical}
In this case, $f$ undergoes a transcritical bifurcation at $(\eta^*, \alpha^*)$.  In a neighbourhood of this point, $f$ satisfies conditions \eqref{eq:non_hyperbolic_equ_1}--\eqref{eq:non_hyperbolic_equ_2} above and we further assume the existence of a smooth branch of equilibria $\eta = \chi(\alpha)$ passing through the bifurcation point, so that
\begin{equation}
    f(\chi(\alpha), \alpha) = 0, \qquad \text{for all } \alpha.
\end{equation}
Differentiating this identity repeatedly with respect to $\alpha$ and evaluating at 
$\alpha = \alpha^*$ yields:
\begin{align}
    \partial_\alpha f(\eta^*,\alpha^*) &= 0, \\
    \partial_{\alpha \alpha}f(\eta^*,\alpha^*) &= -(\chi'(\alpha^*))^2\partial_{\eta \eta}f(\eta^*,\alpha^*)-2\chi'(\alpha^*)\partial_{\alpha \eta}f(\eta^*,\alpha^*). \label{eq:cond_second_order_trans}
\end{align}

Restricting ourselves to generic cases, we assume that $\partial_{\alpha\alpha}f,\, \partial_{\alpha\eta}f,\, \partial_{\eta\eta}f \neq 0$ at $(\eta^*, \alpha^*)$. 
Expanding $f$ in a Taylor series around this point then yields the correspondence:
\begin{align}
    \frac{\mathrm{d}\eta}{\mathrm{d} t} &= \frac{(\alpha-\alpha^*)^2}{2}\partial_{\alpha \alpha} f(\eta^*,\alpha^*)  + (\eta-\eta^*) (\alpha-\alpha^*)\partial_{\alpha \eta} f(\eta^*,\alpha^*) + \frac{(\eta-\eta^*)^2}{2}  \partial_{ \eta \eta} f(\eta^*,\alpha^*)  \nonumber
    \\ & \quad + O\big((\alpha-\alpha^*)^3,\,(\alpha-\alpha^*)^2(\eta-\eta^*),(\alpha-\alpha^*)(\eta-\eta^*)^2,\,(\eta-\eta^*)^3 ), \label{eq:transcritical_DL}
\end{align}
which, for $\alpha-\alpha^*$ and $\eta-\eta^*$ sufficiently small, can be approximated by:
\begin{align}
\frac{\mathrm{d}\eta}{\mathrm{d} t} \approx \frac{(\alpha-\alpha^*)^2}{2}\partial_{\alpha \alpha} f(\eta^*,\alpha^*)  + (\eta-\eta^*) (\alpha-\alpha^*)\partial_{\alpha \eta} f(\eta^*,\alpha^*) + \frac{(\eta-\eta^*)^2}{2}  \partial_{ \eta \eta} f(\eta^*,\alpha^*) . \label{eq:transcritical_approx}
\end{align}

Note that equation~\eqref{eq:transcritical_approx} is not directly similar to the standard normal form of the transcritical bifurcation ($\frac{\mathrm d x}{\mathrm{d} t} = rx - x^2$), as it contains an additional term in $(\alpha-\alpha^*)^2$ (i.e. $r^2$). It is however, possible to find a transformation which  leads to the normal form. Since the right-hand side of equation~\eqref{eq:transcritical_approx} is a degree-two polynomial in $\eta - \eta^*$, computing its discriminant $\Delta$ yields:
\begin{align}
    \Delta &= (\alpha-\alpha^*)^2 \left( (\partial_{\alpha \eta}f(\eta^*,\alpha^*))^2 -\partial_{\alpha \alpha}f(\eta^*,\alpha^*) \partial_{\eta \eta}f(\eta^*,\alpha^*) \right) \\
    &= (\alpha-\alpha^*)^2\left(\partial_{\alpha \eta}f(\eta^*,\alpha^*)  + \chi'(\alpha^*) \partial_{\eta \eta}f(\eta^*,\alpha^*) \right)^2 \\
    &\ge0
\end{align}
where we have used equality ~\eqref{eq:cond_second_order_trans}. We further assume that $\Delta \neq 0$, so as to avoid degenerate cases. Consequently, equation~\eqref{eq:transcritical_approx} admits two distinct real roots of $\eta - \eta^*$ for each value of $\alpha$,
\begin{equation}\label{eq:expression_sigma}
\sigma_\pm(\alpha) =\frac{-\partial_{\alpha \eta} f( \eta^*, \alpha^*) \pm \lvert \partial_{\alpha \eta} f( \eta^*, \alpha^*) + \chi'(\alpha^*)\partial_{\eta \eta} f( \eta^*, \alpha^*)\rvert}{\partial_{\eta \eta} f( \eta^*, \alpha^*)}  (\alpha-\alpha^*),
\end{equation}
and equation~\eqref{eq:transcritical_approx} can then be written as follows:
\begin{equation}\label{eq:approximate_transcritical_reduced}
    \frac{\mathrm{d}\eta}{\mathrm{d} t} \approx  \frac{\partial_{ \eta \eta} f(\eta^*,\alpha^*)}{2}\left((\eta-\eta^*)-\sigma_+(\alpha) \right)\left((\eta-\eta^*)-\sigma_-(\alpha) \right)
\end{equation}

To recover the normal form of the transcritical bifurcation, one can perform a change of variables of the form:
\begin{align}
    x &=\frac{-\partial_{\eta \eta}  f(\eta^*,\alpha^*)}{2}\left( (\eta-\eta^*) - \sigma_+(\alpha) \right) ,\\
    r &= \frac{\partial_{\eta \eta} f( \eta^*, \alpha^*)}{2}( \sigma_+(\alpha)- \sigma_-(\alpha)) = \lvert \partial_{\alpha \eta} f( \eta^*, \alpha^*) + \chi'(\alpha^*) \partial_{\eta \eta} f( \eta^*, \alpha^*)\rvert (\alpha-\alpha^*),
\end{align}

Here, contrary to the fold and pitchfork bifurcation cases, $x$ depends on both $\eta$ and $\alpha$ and therefore this transformation does not lead to a clear conclusion on the resilience metrics. Nevertheless, one can explicitly compute the dependence of the resilience metrics on the distance to the bifurcation point for the approximate model of equation~\eqref{eq:transcritical_approx} (see Appendix~\ref{an:scalings_transcritical}). This recovers scaling laws with the same dependence on $\lvert \alpha - \alpha^* \rvert$ as in Table~\ref{tab:resilience_metrics}:
\begin{equation}
    \lvert \lambda\rvert \propto \lvert \alpha - \alpha^* \rvert \quad ; \quad h \propto \lvert \alpha - \alpha^* \rvert^3 \quad ; \quad ds \propto \lvert \alpha - \alpha^* \rvert
\end{equation}

This shows that the additional term in equation~\eqref{eq:transcritical_approx} in comparison to the normal form, does not qualitatively alter the scalings found in Table \ref{tab:resilience_metrics} near the bifurcation point.

\subsection{Relative dependencies of resilience metrics on the distance to the bifurcation point}
We conclude that, for the three types of local bifurcations considered, power-law dependencies of the resilience metrics on the distance to the bifurcation point are expected to be the same in general as in the normal form framework, at least sufficiently close to the bifurcation point. Since we are only interested in the scaling exponents (i.e., the exponents of $\lvert \alpha - \alpha^* \rvert $) rather than the absolute values of the metrics, the prefactors need not be determined, and the scaling laws derived for the normal form models remain valid. The relationship between general and normal form models will be discussed more thoroughly in section~\ref{sec:discussion}.

\section{Scaling laws validation for specific models}\label{sec:examples_validation}
We verify that the scaling laws derived for the normal form models are recovered in three specific but generally used models, at least sufficiently close to the bifurcation point. We perform this verification using three classical models: an ecological model exhibiting bistability (fold bifurcation), a mixture of Gaussians model (fold and pitchfork bifurcations), and an SIS compartmental model (transcritical bifurcation). The three models are of the form $\frac{\mathrm{d}\eta}{\mathrm{d}t} = f(\eta,\alpha) = -\frac{\mathrm{d} V_\alpha}{\mathrm{d} \eta}(\eta)$, with scalar fields and potential functions summarized in Table~\ref{tab:models_num}.

\begin{table}[h] \centering
\centerline{
\begin{tabular}{c|c|c|c}
        & Bistable ecological model & Mixture of Gaussians model & SIS model  \\ \hline
$f(\eta,\alpha)$  & $\beta \eta(1-\frac{\eta}{K}) -\alpha\frac{\eta^2}{\eta^2+h^2}$ & $- \frac{\eta}{2}e^{- \eta^2/2}- \frac{\eta-\alpha}{2 \sigma^2} e^{-(\eta-\alpha)^2/2\sigma^2}$  & $\beta \eta \left( 1-\frac{1}{\alpha} -\eta\right)$ \\
$V_\alpha(\eta) $  & $-\beta \eta^2 \left(\frac{1}{2} -\frac{\eta}{3 K} \right) +\alpha \left(\eta-h \arctan\left(\frac{\eta}{h} \right) \right)$ & $1- \frac{1}{2}e^{- \eta^2/2}- \frac{1}{2} e^{-(\eta-\alpha)^2/2\sigma^2}$ & $-\beta \eta^2 \left( \frac{1}{2} \left(1-\frac{1}{\alpha}\right) - \frac{\eta}{3}\right)$
\end{tabular}}
\caption{Scalar fields $f$ and potential functions $V_\alpha$ for the three canonical models considered in section~\ref{sec:examples_validation}. The state variable is denoted $\eta$ and the bifurcation parameter is denoted $\alpha$. Other parameters are specified in sections \ref{sec:bistable_eco_model}--\ref{sec:SIS_model}.}\label{tab:models_num}
\end{table}

\subsection{Bistable ecological model}\label{sec:bistable_eco_model}
The first model, hereafter referred to as the bistable ecological model, describes the transition in the population dynamics of a harvested resource toward overexploitation. It is a classical example used to illustrate bistability and hysteresis in ecological systems \cite{noy1975stability, may1977thresholds}, and is therefore widely employed in the EWS literature (see, for instance, \cite{dakos2012methods,dakos2012robustness, dakos2015resilience}). The model describes the dynamics of a resource biomass $\eta$ which grows logistically and is subject to harvesting at a rate $\alpha$. Further parameters are the intrinsic growth rate $\beta$, the carrying capacity $K$, and the half-saturation constant $h$ (see Table~\ref{tab:models_num} for the explicit model).

\begin{figure}[h]
\begin{subfigure}{.49\textwidth}
  \centering
  \includegraphics[width=1.\linewidth]{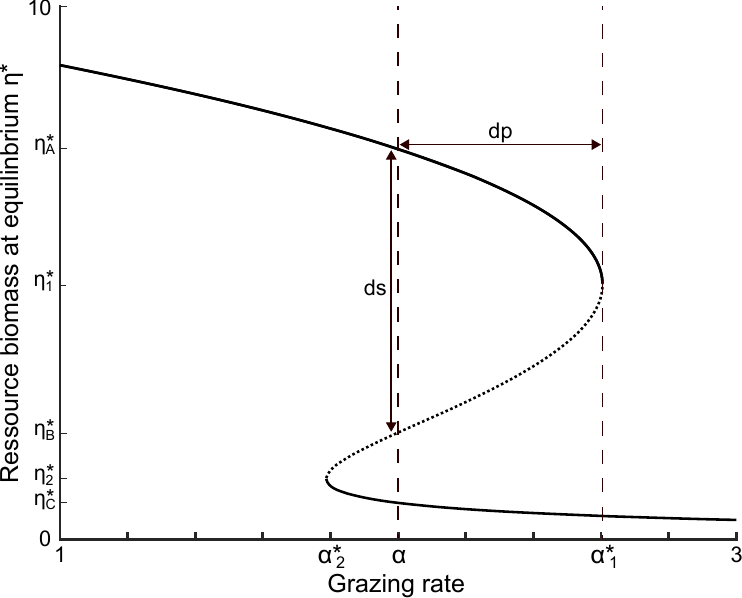}
  \caption{}\label{fig:fig3_bif_diag}
\end{subfigure}%
\begin{subfigure}{.49\textwidth}
  \centering 
  \includegraphics[width=1.\linewidth]{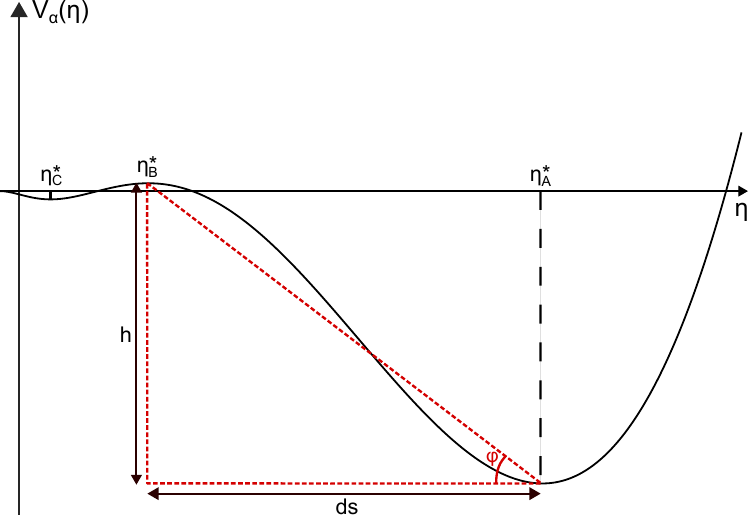}
   \begin{minipage}{.1cm}
            \vfill
            \end{minipage}
  \caption{}\label{fig:fig3_potential}
\end{subfigure}
\caption{Graphical illustration of resilience metrics and their relationship with the distance to the bifurcation point.
Figure~\ref{fig:fig3_bif_diag}: Bifurcation diagram of a classical ecological model of a logistically growing resource under harvesting, studied in section \ref{sec:bistable_eco_model} and considered in \cite{dakos2012methods, dakos2012robustness, dakos2015resilience}. Resource biomass equilibria, denoted $\eta^*$, are plotted as a function of the grazing rate $\alpha$. Two fold bifurcations occur along the curves of equilibria, the first at $(\alpha^*_1,\eta^*_1) \approx (2.60,4.78) $ and a second one at $(\alpha^*_2,\eta^*_2) \approx (1.79, 1.14)$. The figure includes a graphical representation of the basin width, $ds$ (ecological resilience metric), and the distance in parameter space to the critical threshold at which the bifurcation occurs, $dp$ (=$\lvert\alpha-\alpha^*_1\rvert$).
Figure~\ref{fig:fig3_potential}: Associated potential function for a fixed value of $\alpha$, with a representation of the ecological resilience metrics $ds$ and $h$. We also add the indirect resilience metrics $\varphi$ introduced in \cite{dakos2015resilience}.
For this value of $\alpha$, the system admits the trivial equilibrium where the resource is absent and three non-trivial equilibria: $\eta^*_A$ (stable), $\eta^*_B$ (unstable), and $\eta^*_C$ (stable).}\label{fig:correction_dakos2015} 
\end{figure}

Its main interest lies in the fact that, for certain parameter values, the system exhibits bistability. For instance, for the parameter set considered in \cite{dakos2012methods} (namely $\beta = h = 1$ and $K = 10$), the corresponding bifurcation diagram is shown in Figure~\ref{fig:fig3_bif_diag}. Depending on the value of the parameter $\alpha$, the system admits either two or four equilibria, with half of them being locally stable and the other two unstable.

Two fold bifurcations happen along the line of equilibria. We first examine the behaviour of the system in the vicinity of the bifurcation at the larger value of $\alpha$ among the two. The values of the state variable and the bifurcation parameter at this point are denoted $(\eta^*_1, \alpha^*_1)$ (shown in Figure~\ref{fig:fig3_bif_diag}). There is no simple closed-form expression for these values.

One can check that the genericity conditions for the fold, equations~\eqref{eq:non_hyperbolic_equ_3}--\eqref{eq:non_hyperbolic_equ_4}, are satisfied:
\begin{align}
    \partial_{\alpha}f(\eta^*_1,\alpha_1^*) &= - \frac{(\eta^*_1)^2}{(\eta^*_1)^2 + h^2} \approx -0.9581 \neq 0,\\
     \partial_{\eta \eta}f(\eta^*_1,\alpha_1^*) &= \frac{\alpha_1^* h^2(6(\eta^*_1)^2-2h^2)}{((\eta^*_1)^2 + h^2)^3}  - \frac{2\beta}{K} \approx -0.1741 \neq 0,
\end{align}
inserting the parameter values of \cite{dakos2012methods} and using approximations of the bifurcation point obtained with MatCont \cite{dhooge2003matcont}. Similar non-zero conditions hold at the fold bifurcation at the lower value of $\alpha$, namely $(\eta^*_2, \alpha_2^*)$.

To compute the resilience metrics, we apply equations \eqref{eq:dom_ev}, \eqref{eq:barrier_pot}, and \eqref{eq:width_basin}. Yet, the scalings must be validated numerically for this model, as closed-form analytical expressions for the equilibria cannot be derived. The stable and unstable equilibria are computed numerically using Python.

\begin{figure}[h]
  \centering
  \includegraphics[width=.6\linewidth]{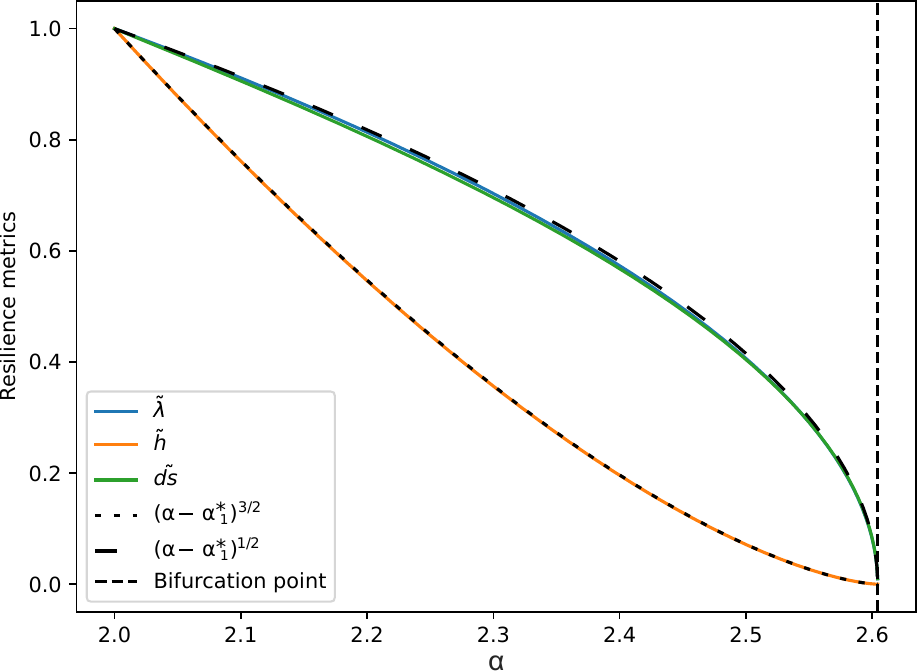} 
\caption{Numerical validation of the scaling laws governing the dependence of resilience metrics on the bifurcation parameter, along the branch of stable equilibria near the upper fold bifurcation at $ \alpha =  \alpha^*_1 \approx 2.60$.
Shown is the empirical dependence of three resilience metrics ($\lambda$, $h$, and $ds$) on the bifurcation parameter $\alpha$ for the ecological model studied in section~\ref{sec:bistable_eco_model}, over the range $\alpha \in [2.0,  \alpha^*_1]$, corresponding to the approach to the upper fold bifurcation.
Values of the metrics are normalized by their respective values at $ \alpha = 2.0$ and denoted $\tilde{\lambda}$, $\tilde{h}$, and $\tilde{ds}$, so as to bring them to comparable orders of magnitude and enable a meaningful comparison of their decay rates within the interval shown. 
Theoretical scalings predicted by the fold normal form and reported in Table~\ref{tab:resilience_metrics} are shown as dotted or striped black lines.}\label{fig:fig4_fold_dependances_resilience} 
\end{figure}

Figure~\ref{fig:fig4_fold_dependances_resilience} presents the values of $\lambda$, $ds$, and $h$ as functions of the bifurcation parameter $\alpha$, along the branch of stable equilibria (upper branch) leading to the bifurcation point at $\alpha^*_1$. Values of $\alpha$ are chosen in the interval $(2.0,\alpha^*_1)$, where two stable equilibria coexist and the ecological resilience metrics are therefore well defined. To facilitate comparison within the same figure, all metrics are normalized by their value at $\alpha = 2.0$, so that their relative decay rates within the interval can be compared directly. This is consistent with our focus on the power-law dependence on the distance to the bifurcation point, rather than on the absolute values of the resilience metrics.

We observe that the dependence on $\lvert \alpha-\alpha^*_1 \rvert$ differs across the three metrics. The scalings of $\lambda$ and $ds$ are very close to an $\lvert \alpha  - \alpha^*_1 \rvert^{1/2}$ scaling, while $h$ decays as $\lvert \alpha-\alpha^*_1 \rvert^{3/2}$, both consistent with the fold normal form predictions of section~\ref{sec:links_eco_eng_resilience}. Since the theoretical predictions hold far from the bifurcation---the interval of grazing rates considered here spans nearly the entire bistable region, see Figure~\ref{fig:fig3_bif_diag}---we now turn to the lower fold bifurcation, where the neighbourhood of validity is smaller.

\begin{figure}[h!]
\centering
    \begin{subfigure}{.48\textwidth}
  \includegraphics[width=1.\linewidth]{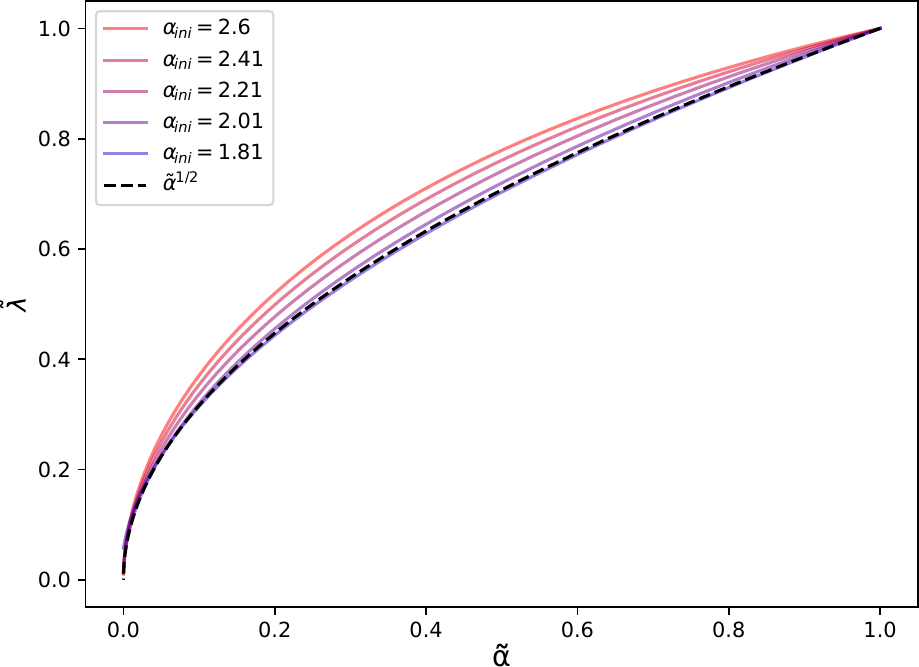}
  \caption{}\label{fig:fig5_lambda}
\end{subfigure} 
\begin{subfigure}{.48\textwidth}
  \includegraphics[width=1.\linewidth]{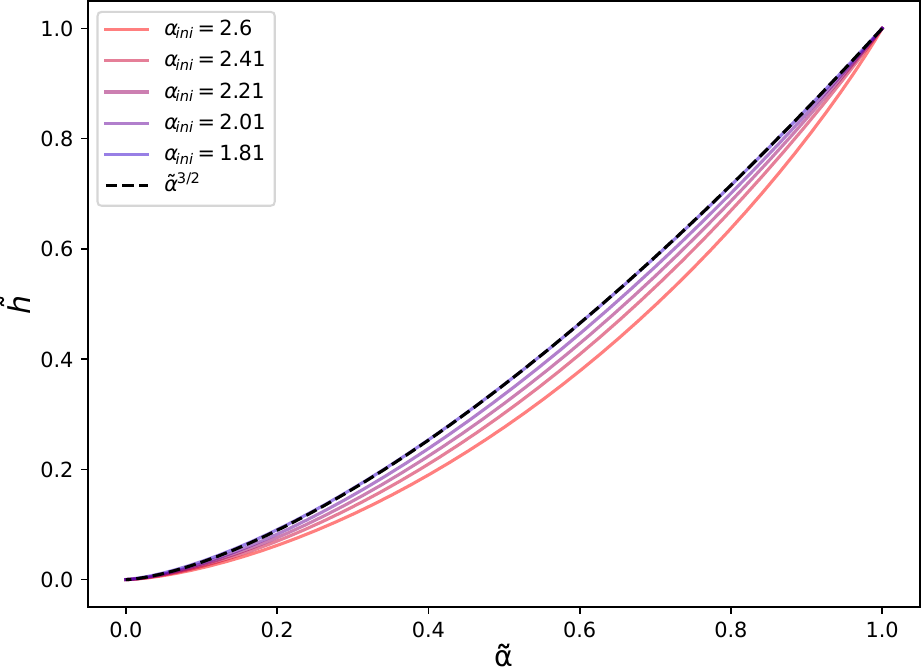}
  \caption{}\label{fig:fig5_h}
\end{subfigure}
\begin{subfigure}{.48\textwidth}
  \includegraphics[width=1.\linewidth]{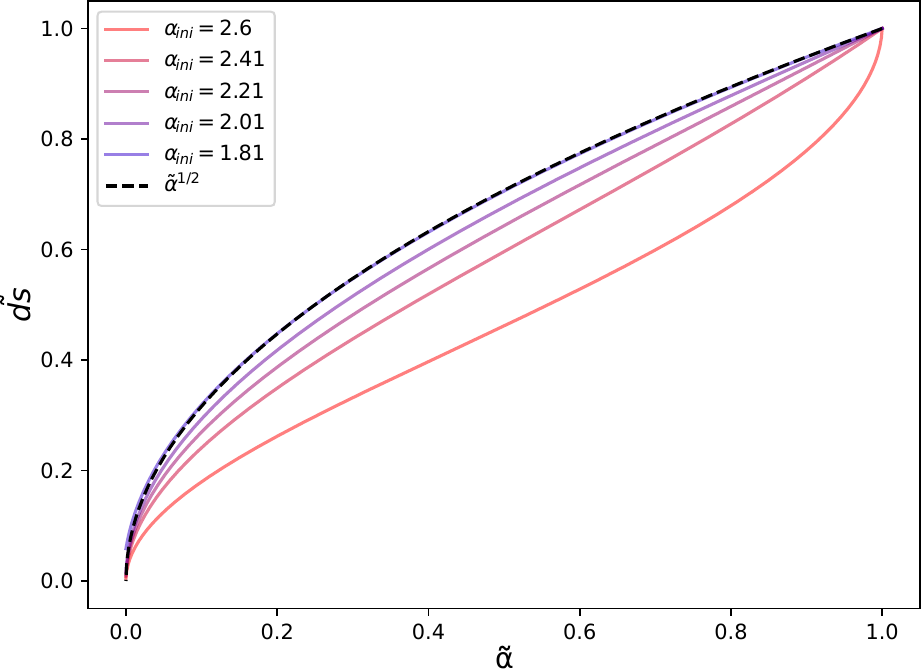}
  \caption{}\label{fig:fig5_ds}
\end{subfigure}
\begin{subfigure}{.48\textwidth}
  \includegraphics[width=1.\linewidth]{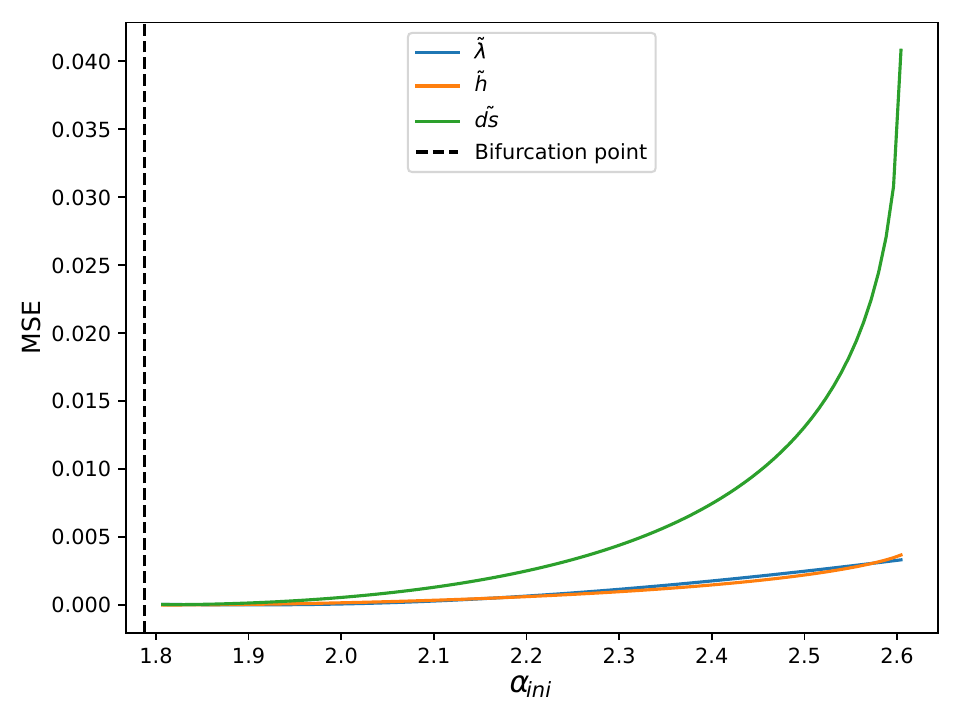}
  \caption{}\label{fig:fig5_MSE}
\end{subfigure}
    \caption{Numerical comparison of the decay rates of resilience metrics as a function of the distance to the bifurcation point for the bistable ecological model studied in section~\ref{sec:bistable_eco_model}, near the lower fold bifurcation at $\alpha = \alpha^*_2 \approx 1.79$. Figures~\ref{fig:fig5_lambda}--\ref{fig:fig5_ds} show the resilience metrics over $\alpha \in [\alpha_{\mathrm{ini}}, \alpha^*_2]$ (using $100$ evenly spaced values of $\alpha$), normalized by their values at $\alpha = \alpha_{\mathrm{ini}}$, for several choices of $\alpha_{\mathrm{ini}}$. The horizontal axis is the normalized distance to the bifurcation point $\tilde{\alpha} = (\alpha - \alpha_{\mathrm{ini}})/(\alpha^*_2 - \alpha_{\mathrm{ini}})$. Dotted lines indicate the theoretical scalings from Table~\ref{tab:resilience_metrics}.  Figure~\ref{fig:fig5_MSE} summarizes this result by reporting the mean squared error (MSE) over the evenly spaced values between the empirical resilience curves shown in Figures \ref{fig:fig5_lambda}–\ref{fig:fig5_ds} (solid lines) and the corresponding theoretical predictions (dotted lines), as a function of $ \alpha_{\mathrm{ini}}$. As $\alpha_{\mathrm{ini}}$ approaches the bifurcation value $\alpha^*_2$, the MSE decreases, indicating that the scaling relationships predicted by the normal form become progressively more accurate.}
    \label{fig:fig5_distance_bifurcation} 
\end{figure}

Figure~\ref{fig:fig5_distance_bifurcation} compares the empirical decay rates of the resilience metrics with the theoretical predictions of Table~\ref{tab:resilience_metrics}, over neighbourhoods of varying sizes around the lower fold bifurcation at $\alpha^*_2$. When the distances within the interval from the bifurcation point decrease, the agreement between the numerical results and the normal-form predictions increases, as further quantified by the mean squared error (MSE) reported in Figure~\ref{fig:fig5_MSE}. 

The fact that the scaling laws hold over larger neighbourhoods for the upper fold bifurcation than for the lower one highlights that the extent of the neighbourhood over which the scaling laws remain valid is inherently specific to the bifurcation point under consideration.

\subsection{Mixture of Gaussians model}\label{sec:potential_gaussian}
In this section, we consider a potential which is a mixture of Gaussian functions, given in Table~\ref{tab:models_num}. It is similar to the potential considered by Dakos and Kéfi \cite{dakos2022ecological} (Equation~(B.2)) and to the one studied by Nolting and Abbott \cite{nolting2016balls} (Equation~(S40)), here restricted to one dimension. We investigate how resilience metrics behave as the system undergoes a qualitative change in stability, by varying the separation between the peaks of the two Gaussians through the parameter $\alpha \geq 0$. Depending on the value of $\sigma$---which controls the asymmetry between the two Gaussians---different types of bifurcations may arise, allowing us to tune the nature of the transition.

\begin{figure}
\begin{subfigure}{.45\textwidth}
  \centering
  \includegraphics[width=1.\linewidth]{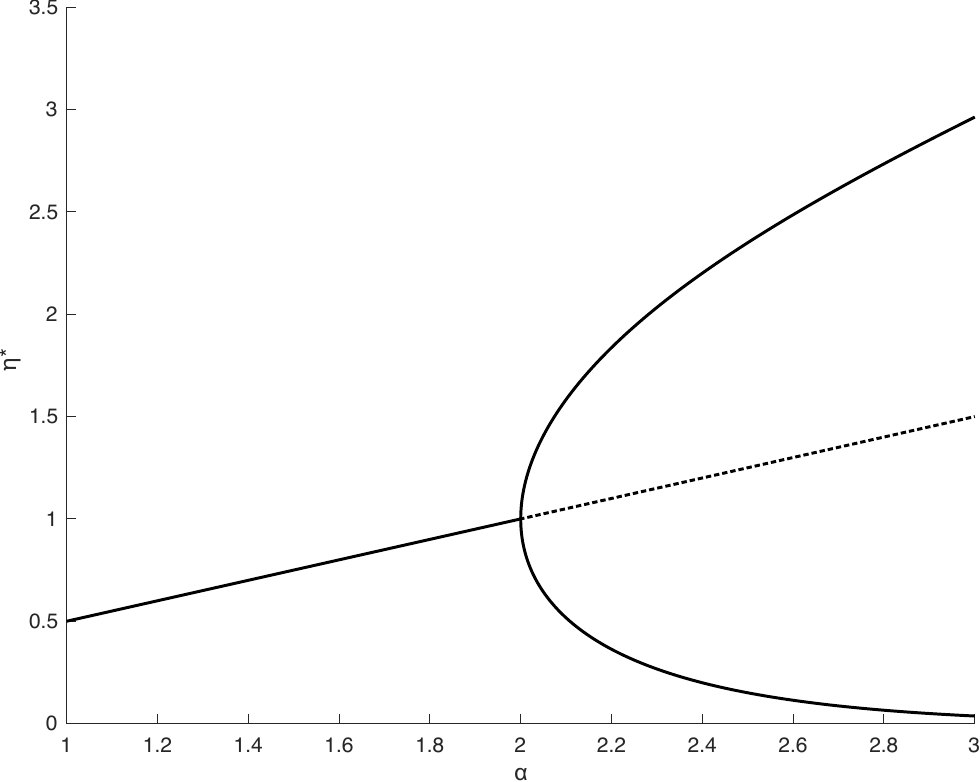}
  \caption{$\sigma=1.0$ (pitchfork bifurcation)}\label{fig:an_sym_gaussian_bif_diag}
\end{subfigure}%
\hspace{1cm} 
\begin{subfigure}{.45\textwidth}
  \centering
  \includegraphics[width=1.\linewidth]{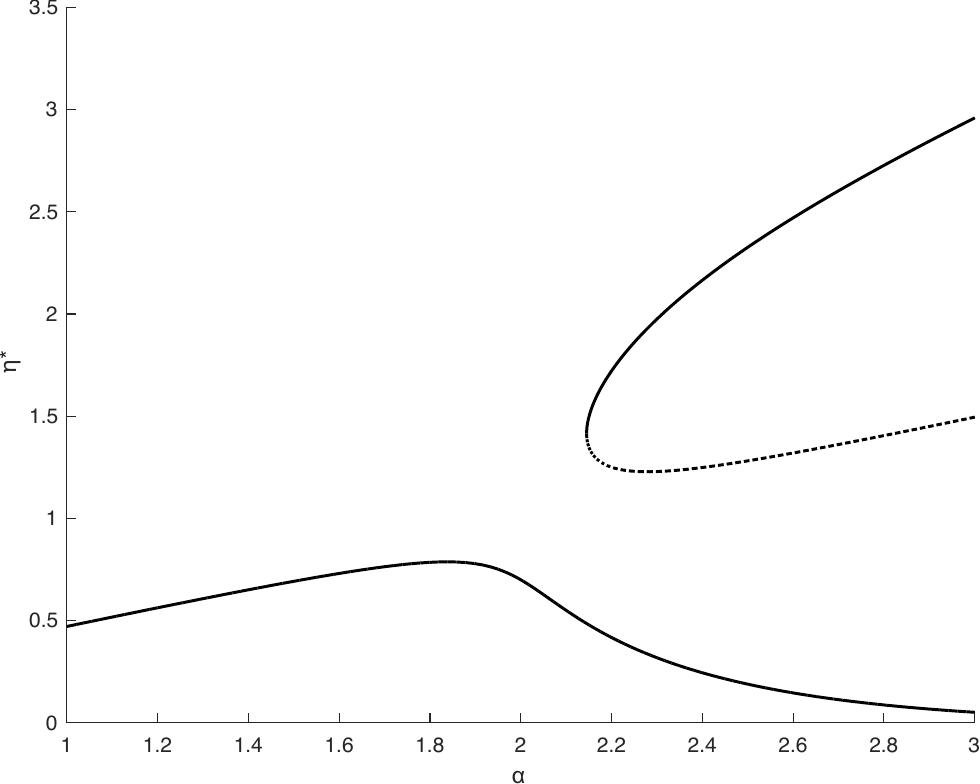}
  \caption{$\sigma=1.05$ (fold bifurcation)}\label{fig:an_asym_gaussian_bif_diag}
\end{subfigure}
\caption{Bifurcation diagram of the Gaussian potential defined in Table~\ref{tab:models_num}, illustrating the effect of the asymmetry parameter $\sigma$.}
\end{figure}

When $\sigma = 1$, the Gaussian potential corresponds to a mixture of two identical Gaussian distributions with unit standard deviation. In this case, the resulting function is bimodal for $ \alpha \ge 2$ and unimodal otherwise \cite{sitek2016modes}. Intuitively, when the two Gaussians are sufficiently close, they merge into a single peak. This result does not generally hold when $\sigma \neq 1$. Consequently, for $\sigma = 1$, we expect the potential to transition from a double-well to a single-well structure as $\alpha$ crosses the critical value $\alpha^* = 2$, leading to a supercritical pitchfork bifurcation. The corresponding bifurcation diagram is shown in Figure~\ref{fig:an_sym_gaussian_bif_diag}.

When $\sigma \neq 1$, the asymmetry between the two Gaussian components prevents the occurrence of a supercritical pitchfork bifurcation. Instead, the system undergoes a fold bifurcation at a critical value of $\alpha$ (see Figure~\ref{fig:an_asym_gaussian_bif_diag}). This behaviour reflects the generic nature of fold bifurcations, in contrast to pitchfork bifurcations, which are structurally unstable and require specific symmetry conditions. As a result, even small parameter perturbations---such as the introduction of asymmetry in the present case---break the symmetry underlying the pitchfork bifurcation and unfold it into a fold bifurcation.

To compute the resilience metrics, we apply equations \eqref{eq:dom_ev}, \eqref{eq:barrier_pot}, and \eqref{eq:width_basin}, with stable and unstable equilibria computed numerically using Python.

\begin{figure}[h]
\centering
  \includegraphics[width=1.\linewidth]{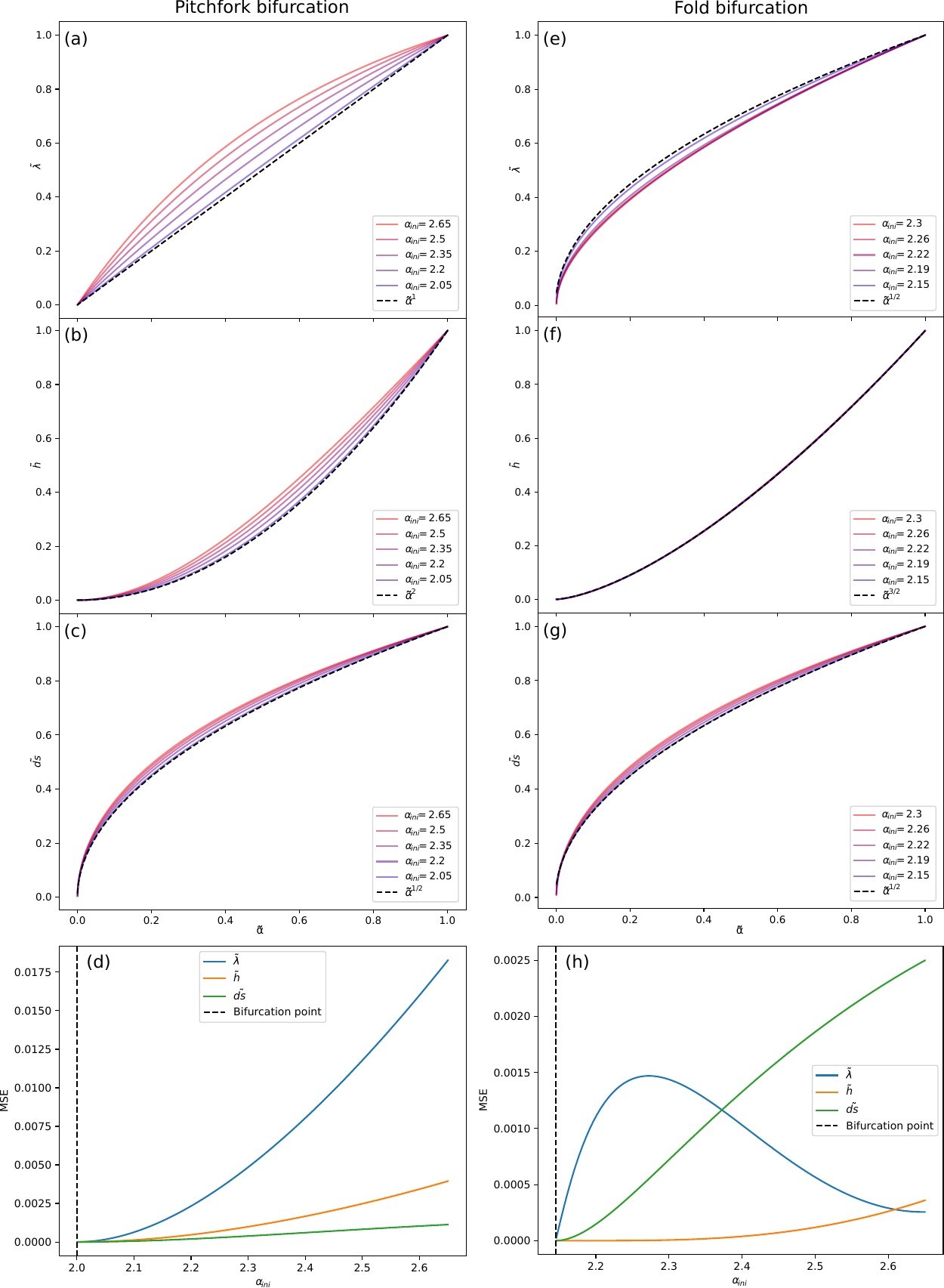}
    \caption{Same Figure as Figure~\ref{fig:fig5_distance_bifurcation} but for the mixture of Gaussians potential of Table~\ref{tab:models_num}. Sub-figures (a)--(d) are the symmetric case ($\sigma = 1$, bifurcation diagram on Figure~\ref{fig:an_sym_gaussian_bif_diag}, with a pitchfork bifurcation), near the upper stable branch of the pitchfork bifurcation point at $\alpha_c = 2$. Sub-figures (e)--(h) are the asymmetric case ($\sigma = 1.05$, bifurcation diagram on Figure~\ref{fig:an_asym_gaussian_bif_diag}, with a fold bifurcation), near the upper stable branch of the fold bifurcation point at $\alpha_c \neq 2$. Resilience metrics denoted with a tilde are normalized by their values at $\alpha = \alpha_{\mathrm{ini}}$, while the horizontal axis corresponds to the normalized distance to the bifurcation point, defined as $\tilde \alpha = \frac{\alpha - \alpha_{\mathrm{ini}}}{\alpha_c - \alpha_{\mathrm{ini}}}$.}
    \label{fig:an_gaussian_scalings} 
\end{figure}

Figure~\ref{fig:an_gaussian_scalings} presents the decay rates of the main resilience metrics as functions of the distance to the bifurcation point, together with the associated error rates measured by the mean squared error, for the two types of bifurcation ($\sigma = 1$ or $\sigma \neq 1$ cases) and the three resilience indicators. We observe that the different scalings predicted for fold and supercritical pitchfork bifurcations are progressively recovered as the system approaches the bifurcation point. 
As best illustrated in Figure~\ref{fig:an_gaussian_scalings}h, the scaling of $\lambda$ first moves away from the prediction, the agreement being recovered only sufficiently close to the bifurcation point.

Notably, theoretical scalings differ for the $\lambda$ and $h$ indicators between fold and supercritical pitchfork bifurcations, and both regimes are recovered within the same model by varying the parameter $\sigma$, as expected on the basis of our results for normal forms.

\subsection{SIS compartmental model}\label{sec:SIS_model}
Another useful example is the classical one-dimensional SIS (Susceptible–Infected–Susceptible) compartmental model, modelling the prevalence of an infectious disease over time. $\eta$ is the density of infected individuals at time $t$, $\alpha = \beta/\gamma$ is the basic reproduction number of the model (often denoted $R_0$) where $\beta$ is the transmission rate and $\gamma$ the recovery rate of the disease. In the following, we assume $\alpha>0$, which is a restriction to the biologically relevant cases. The equation governing changes over time in $\eta$ is given in Table~\ref{tab:models_num}.

The system has one or two fixed points depending on the value of $\alpha$: the disease-free state ($\eta^*=0$) which is stable whenever $\alpha<1$ and unstable whenever $\alpha>1$ and the endemic state ($\eta^* = 1-\alpha^{-1}$) is positive and stable whenever $\alpha>1$. Therefore, $\alpha$ plays the role of a bifurcation parameter for a transcritical bifurcation occurring at $\alpha^* = 1$.

Unlike the models considered above, this model admits a fully analytical treatment. At the bifurcation point, the following conditions on $f$ hold true:
\begin{align}
    f(\eta^*,\alpha^*)&=0\\ 
    \partial_\eta f(\eta^*,\alpha^*)&=0,\\
    \partial_\alpha f(\eta^*,\alpha^*)&=0, \\
    \partial_{\eta \eta} f(\eta^*,\alpha^*) &= -2 \beta,\\
    \partial_{\eta \alpha} f(\eta^*,\alpha^*) &=\partial_{\alpha \eta} f(\eta^*,\alpha^*) = \beta,\\
    \partial_{\alpha \alpha} f(\eta^*,\alpha^*) &= 0.
\end{align}

Note that $\partial_{\alpha\alpha} f(\eta^*,\alpha^*) = 0$ implies that this is a degenerate case of section~\ref{sec:normal_form_reduction_transcritical}, where we assumed $\partial_{\alpha\alpha} f(\eta^*,\alpha^*) \neq 0$ for genericity. Here, however, $\partial_{\alpha^i} f(\eta^*,\alpha^*) = 0$ for all $i$, so that the term $\partial_{\alpha\alpha} f$ vanishes in equation~\eqref{eq:transcritical_approx}, without the need to carry the expansion to higher order.

Since explicit expressions for $V$ and $f'$ are available, the only quantities required are the stable and unstable equilibria, denoted $\eta^*_{\mathrm{stab}}$ and $\eta^*_{\mathrm{unstab}}$, which are also known analytically as shown above. This allows us to derive closed-form expressions for the resilience metrics at any value of $\alpha$:
\begin{align}
    \lambda &=  -\frac{\beta}{\alpha} \left\lvert \alpha-1 \right\rvert \underset{\alpha\to1}{\sim} - \beta\lvert \alpha-1 \rvert, \\
    ds &= \frac{\left\lvert \alpha-1\right\rvert}{\alpha} \underset{\alpha\to1}{\sim} \lvert \alpha-1 \rvert, \\
    h &= \frac{\beta}{6\alpha^3} \left\lvert \alpha-1 \right\rvert^3 \underset{\alpha\to1}{\sim} \frac{\beta}{6} \lvert \alpha-1 \rvert^3,
\end{align}
and thus recover the local relative rates at which the different resilience metrics decay to zero depending on the distance to the bifurcation point ($\lvert \alpha-1\rvert$). These are scalings with power laws predicted by Table \ref{tab:resilience_metrics} for the transcritical bifurcation normal form, and equal to them up to a scaling factor.

\section{Comparison with previously proposed scaling laws}\label{sec:comparison_scalings}
Two attempts to link engineering and ecological resilience metrics under bifurcation-induced tipping points were previously proposed: in Box~1 of \cite{dakos2015resilience} and in Supplementary Text~S2 of \cite{dai2015relation}. While the latter recovers the same scalings as ours, it is restricted to the fold bifurcation and does not prove whether these normal form results extend beyond these models. In the former, the scalings proposed---also limited to the fold bifurcation and without formal derivation---are inconsistent with those obtained from our normal form analysis. A summary of the scalings and relationships between resilience metrics found in these two publications is provided in Table~\ref{tab:table_scalings_literature}, which we compare with our results from Table~\ref{tab:resilience_metrics}. Figure~\ref{fig:correction_dakos2015} provides a graphical representation of resilience metrics for the fold bifurcation. We also recall the angle $\varphi$ introduced in \cite{dakos2015resilience} and shown in Figure~\ref{fig:fig3_potential}, from which one obtains
\begin{equation}
    \tan(\varphi) = \frac{h}{ds}.
\end{equation}

\begin{table}[h]
    \centering
    \begin{tabular}{c|c|c}
          & Dakos et al. (2015) \cite{dakos2015resilience} & Dai et al. (2015) \cite{dai2015relation} \\ \hline
         $\lvert \lambda \rvert $ & $\approx \frac{h}{ds}$ & $\propto  dp^{1/2} $ \\
         $ h $ & $\approx dp$ & --- \\
         $ds$ & --- & $\propto dp^{1/2} $ \\
             $ \varphi $ & $\approx \lvert \lambda \rvert $ & ---
    \end{tabular}
    \caption{Relationships between resilience metrics as a function of the distance to the bifurcation point, denoted $dp$, for a fold bifurcation, as reported in \cite{dakos2015resilience} and \cite{dai2015relation}. The angle $\varphi$---an indirect resilience metrics---is as defined in Figure~\ref{fig:fig3_potential}.}
    \label{tab:table_scalings_literature}
\end{table}

The power-law scalings that we derived from normal form models were proved to extend to general models in section~\ref{sec:normal_form} and were numerically validated for model examples in section~\ref{sec:examples_validation}. In the vicinity of a fold bifurcation, we therefore have that $h \propto dp^{3/2}$ where $dp$ is the distance to the bifurcation point in parameter space (denoted $\lvert r \rvert$ in the normal form approach and $\lvert \alpha - \alpha^* \rvert$ for the general models of section~\ref{sec:examples_validation}). As a consequence, $h \not\approx  dp $, contrary to the claim made in \cite{dakos2015resilience}. Although both quantities vanish as $dp\to 0$, they do so with different speeds. 

Furthermore, \cite{dakos2015resilience} implicitly merges the angle $\varphi$ with the absolute value of the eigenvalue $\lambda$. These quantities are not directly related: while $\varphi$ is associated with the geometry of the entire basin, $\lambda$ describes the curvature of the potential at the stable equilibrium and is related to local changes in the tangent rather than its value. This identification is therefore incorrect in general. In Appendix~\ref{an:curvature}, we explicitly relate the angle formed by the tangent at points close to the stable equilibrium to the curvature at that equilibrium, and hence to its eigenvalue.

Using the scalings that we derived for the fold bifurcation, we obtain
\begin{equation}
    \tan(\varphi) = \frac{h}{ds}
    \propto \frac{dp^{3/2}}{{dp}^{1/2}}
    = dp.
\end{equation}

As $\tan(\varphi) \to 0$ when $dp \to 0$, it follows that $\varphi \to 0$, and hence
\begin{equation}
    \varphi \propto dp.
\end{equation}

This approach provides a rigorous derivation of the correct scaling for the angle $\varphi$ near a fold bifurcation. Box~1 of \cite{dakos2015resilience} proposes the relationship $ \varphi \approx |\lambda|$. However, since $|\lambda| \propto dp^{1/2}$, this is inconsistent with the scaling $\varphi \propto dp$. $|\lambda| \not\approx \varphi$ in general.

\section{Discussion}\label{sec:discussion}
We have presented a theoretical framework that clarifies the relationship between ecological and engineering resilience in the context of bifurcation-induced tipping. Although these two notions are often treated as conceptually distinct, our results show that, sufficiently close to a local bifurcation, they are intrinsically linked and can be viewed as complementary descriptions of the same underlying dynamical phenomenon. This contributes to clarifying the relationships between resilience metrics, although only in the vicinity of critical transitions, a challenge that has been identified as central in recent studies \cite{capdevila2021reconciling, dakos2022ecological}. To achieve this, we identified the dominant nonlinearities governing the dynamics near each type of local bifurcation, an approach analogous to studying their normal forms. Starting from a general model exhibiting a local bifurcation, the resilience metrics share the same dependence on the distance to the bifurcation point in parameter space as those of the corresponding normal form model, up to a scaling factor and in the vicinity of the bifurcation point. This approach therefore yields qualitative predictions of the relative behaviour of resilience metrics as the bifurcation point is approached, and allows us to correct previously proposed scalings in the literature that lacked a formal theoretical foundation. Our scalings have been validated both analytically and numerically for three distinct models, covering the three most commonly encountered local bifurcations in one-dimensional systems.

We focused on the most used resilience metrics, but one could derive similar explicit expressions for other resilience metrics \cite{krakovska2024resilience} or even other stability metrics that quantify the way in which a system withstands disturbances \cite{van2021unifying}.

\subsubsection*{Engineering and ecological resilience are equivalent in the vicinity of a local bifurcation}
\begin{figure}[h]
\centering
\includegraphics[width=1.\linewidth]{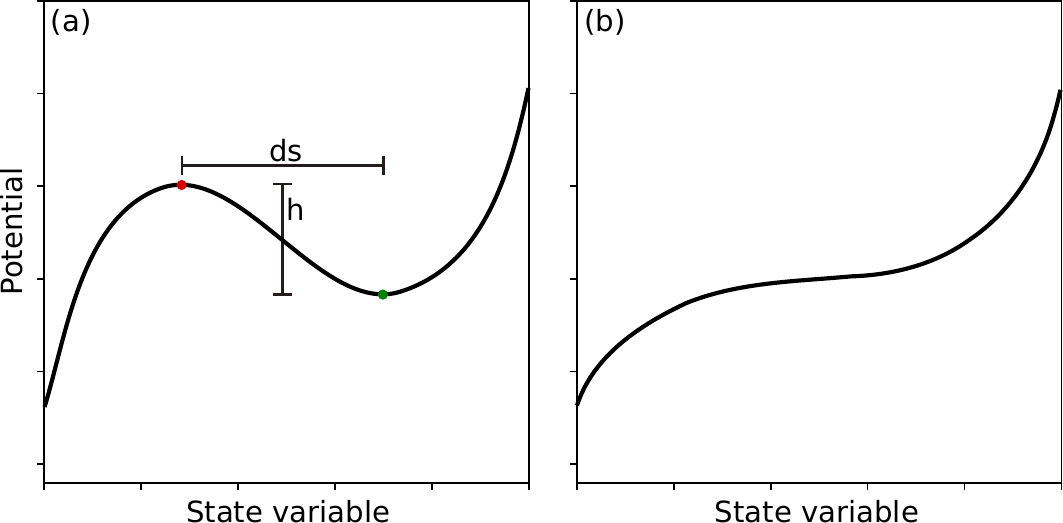}
\caption{Figure~(a) shows the potential function in the neighbourhood of a stable equilibrium (green circle). If the stable equilibrium has disappeared (Figure~(b)) due to a smooth change in the shape of the potential, then both the width of the basin of attraction (quantified here by $ds$, the minimum distance to the basin boundary, represented by the unstable equilibrium in red) and the height $h$ of the potential barrier must necessarily have collapsed to zero.}
\label{fig:transformation_stable_instable} 
\end{figure}

As emphasized recently by Dakos et al. \cite{dakos2022ecological}, theory related to EWS of bifurcation-induced tipping is based on changes in local stability properties which stand "somewhere between ecological and engineering resilience but do not fully bridge the gap". At the same time, previous studies already reported strong correlations between metrics of the two types of resilience in simple ecological models \cite{van2007slow, dakos2015resilience}, although without a theoretical explanation. Our analysis shows that, in the specific setting of bifurcation-induced tipping in one-dimensional systems, such correlations are not coincidental. It provides an intuitive explanation for why ecological and engineering resilience are linked near local bifurcations: for an equilibrium to change stability or disappear, its basin of attraction must necessarily be altered. As illustrated in Figure~\ref{fig:transformation_stable_instable}, the disappearance of a stable equilibrium through a fold bifurcation---which requires the eigenvalue to vanish---cannot occur without the collapse of both the width of the basin of attraction of the equilibrium and the height of its potential barrier. In this sense, basin-level (ecological) and local (engineering) properties are two sides of the same coin when the transition is driven by a local bifurcation. This also explains the strong correlations observed empirically in earlier studies. This suggests that  in real-world systems subject to bifurcation-induced tipping, one could focus an analysis on the type of resilience metric which is easiest to handle, and relationships with other metrics are known in the vicinity of a tipping point. This echoes the original idea of EWS \cite{van2007slow}: find easily measurable indicators that are indirect measures of ecological resilience. Meyer noted that return rates to a stable equilibrium play an important role in determining ecological resilience when they are considered over the entire basin of attraction, and not merely at the equilibrium itself \cite{meyer2016mathematical}. However, sufficiently close to a bifurcation point, the global structure of the basin of attraction is effectively captured by the local curvature at the equilibrium. Note that this need not be the case when the system does not undergo a bifurcation, so that all combinations of engineering and ecological resilience are in principle possible, as illustrated in Figure~\ref{fig:basin_types}.

The fact that resilience metrics are correlated has deeper consequences for stability studies. Van Meerbeek et al.\ \cite{van2021unifying} identify five properties that jointly define stability, among which engineering resilience and ecological resilience. Our results show that these two properties are correlated in the vicinity of a local bifurcation, which effectively reduces the dimensionality of ecological stability in this regime.

\subsubsection*{Explicit scalings for resilience metrics}
Our work provides explicit analytical expressions and scaling laws for commonly used resilience metrics. While normal forms of local bifurcations have been used many times to derive analytical expressions of engineering resilience metrics and therefore EWS in the case of bifurcation-induced tipping \cite{kuehn2011mathematical, o2018stochasticity, bury2020detecting, dylewsky2024early, donovan2024characterising}, they have never been used to derive ecological resilience metrics to the best of our knowledge. Our results make it possible to compare the relative rates at which the different notions of resilience decrease near a local bifurcation. In particular, we show that some metrics typically decay faster than others as the bifurcation point is approached. For example, in the case of a fold bifurcation, the dominant eigenvalue scales as $dp^{1/2}$, while ecological metrics such as the potential barrier height scale as $dp^{3/2}$, where $dp$ is the distance, in parameter space, to the bifurcation point. This difference in scaling exponents implies that relative changes in some resilience metrics might only be detectable at smaller distances from the bifurcation point than others. This suggests that an analysis focusing on a set of metrics can be more informative than focusing on a single one.

Resilience scalings are not universal, but depend on the type of local bifurcation under consideration. Fold, transcritical, and pitchfork bifurcations are governed by different nonlinearities, which directly translate into different scaling behaviours of resilience metrics. This observation reinforces the idea that comparing absolute resilience values between dynamical systems is not relevant, while comparing relative values at different distances from a bifurcation point is. These ideas have already been highlighted in the normal form approach for engineering resilience metrics \cite{kuehn2011mathematical, o2018stochasticity, bury2020detecting}. Furthermore, estimating the scalings associated with critical transitions has been proposed as a promising way to distinguish between potential underlying bifurcations \cite{bury2020detecting}, and our results give predictions for these scalings which help to achieve this.

\subsubsection*{Bifurcation-induced tipping provides a context for comparing resilience metrics}
A crucial issue now becomes predicting when these correlations between resilience metrics may fail to hold. Several attempts to address this have been made in the literature. Menck et al. proposed a one-dimensional model with a discontinuous scalar field, in which the properties of the stochastic state variable changed abruptly depending on its value \cite{menck2013basin}. Such non-smoothness places this case outside the scope of our study, as we focus on "true" bifurcations in the sense of classical dynamical systems theory (see \cite{kuznetsov1998elements}). 

A different study was able to decouple the relationship between engineering resilience and ecology close to a bifurcation, resulting in an increase in the speed of return to equilibrium simultaneously with a decrease in the time of first departure from the basin of attraction \cite{titus2020critical}, a phenomenon called \textit{critical speeding up}. However, as noted by the authors, their model was intended to avoid crossing a bifurcation. They were therefore rather interested in noise-induced tipping while increasing the engineering resilience, rather than allowing bifurcation-induced transitions. Similarly, Dakos and Kéfi \cite{dakos2022ecological} reported weaker correlations between certain resilience metrics using a potential function defined as a tunable mixture of Gaussians, while Nolting and Abbott \cite{nolting2016balls} showed that all possible combinations of three resilience metrics can occur in a similar, but two-dimensional, Gaussian potential. We argue that our results are not in contradiction with their conclusions, as these studies do not consider situations in which the potential transitions from a bimodal to a unimodal structure, a change which would correspond to a bifurcation, but only bistable systems. In section \ref{sec:potential_gaussian}, we demonstrated that when a system defined by such Gaussian potentials undergoes a local bifurcation, the resulting dynamics recover the scaling laws derived in the main text, thereby establishing the consistency of their findings with our theoretical framework.

Other authors looked at scenarios where the positive correlation between engineering (although they call it stability) and ecological resilience does not hold, when two environmental parameters could be changed \cite{dai2015relation}. They found the relationship which we expect when experimental populations were driven sufficiently close towards a bifurcation. It broke down when multiple drivers were changed simultaneously. The authors indicate that this correlation break-down is related to the fact that the system is not approaching the bifurcation sufficiently closely. In this case, they could observe all possible patterns of resilience correlation, including increases in one metric and decreases in another. They also propose a scaling for the minimum distance to the basin boundary (see Table~\ref{tab:table_scalings_literature}), for the fold bifurcation, which is the one which we derived (see Table~\ref{tab:resilience_metrics}).
  
This clearly highlights that the conditions in our study are specific for systems approaching a bifurcation, and that they may no longer be met in more realistic models where correlations between parameters can arise naturally when the system is submitted to an external driver which is not mainly driving them closer to a bifurcation. Scalings proposed in this article are therefore an additional theoretical tool to add to the EWS toolbox, but they are obviously not universal rules which can be applied to any changing dynamical system. We also note that we cannot exclude that scalings with precisely the same properties might arise when changing parameters while a system is not in the vicinity of a bifurcation, but we expect this to occur at most occasionally.

In line with most research on EWS, we focus here on EWS of local bifurcations. Regarding global bifurcations---which involve large regions of phase space and are not confined to the neighbourhood of a single equilibrium---no eigenvalue is required to cross the imaginary axis as the system passes through the bifurcation point, so that no loss of engineering resilience is expected \cite{williamson2015detection, tirabassi2022correlation}. Since EWS of local bifurcations are primarily based on the loss of engineering resilience, they are not expected to detect global bifurcations \cite{hastings2010regime, bel2012gradual, williamson2015detection, tirabassi2022correlation}, although alternative indicators have been proposed to detect such transitions \cite{williamson2015detection, adamson2020forecasting, tirabassi2022correlation}.

Furthermore, our results are obtained for one-dimensional models. While such models are standard tools in resilience studies for theoretical ecology and evolutionary biology, they necessarily simplify the complexity of real ecological systems. As noted previously \cite{dakos2022ecological}, our understanding of the relationship between engineering and ecological resilience may be strongly influenced by the types of models used to study these concepts. Although low-dimensional models lack realism, they nevertheless provide valuable insight into mechanisms that would otherwise remain analytically intractable and their use is justified by the center manifold theorem, which provides a systematic way to reduce the dimensionality of the state space when analysing local bifurcations \cite{guckenheimer1983nonlinear, kuznetsov1998elements}. 

\subsubsection*{To what extent are normal-form models adequate to study local bifurcations?}
Although normal forms are widely used in the EWS literature, evidence that the results derived for them remain valid beyond the normal form setting has generally been lacking. Addressing this question first requires clarifying the distinction between classical normal form theory and the approach adopted here. Normal form theory establishes the existence of smooth changes of variables under which a dynamical system becomes topologically equivalent to a canonical expression known as the normal form (see, e.g., Theorem~3.2 of \cite{kuznetsov1998elements} for a proof of topological equivalence in the case of the fold bifurcation). In contrast, our aim is to simplify the general model by retaining only the nonlinear terms that dominate in the vicinity of the bifurcation point. This may leave more terms than in the normal form---as illustrated by the transcritical case in section~\ref{sec:normal_form_reduction_transcritical}---even though a qualitative equivalence between the two expressions could still be established. For the results sought here, such a further reduction is not required: retaining the full set of dominant nonlinear terms is in fact preferable, as it allows us to derive quantitative scaling laws rather than results of a purely topological nature.

Building on this approach, we have shown in Section~\ref{sec:normal_form} that the resilience scalings derived from normal form models extend to general models exhibiting local bifurcations, up to a scaling factor. The reduced models obtained by retaining only the dominant nonlinear terms indeed yield the same functional dependence on the distance to the bifurcation point as the corresponding normal forms, which provides a formal justification for using normal-form-based scalings in more general settings. Since EWS are aimed at detecting relative trends in resilience metrics, the scaling factor does not affect the dependence on the distance to the bifurcation point; however, if the absolute values of resilience metrics in real systems are of concern, determining this factor becomes essential. The numerical and analytical validation of our scalings across three representative models confirms that normal-form-based resilience metrics can provide meaningful insight into the behaviour of more complex systems near a bifurcation point.

This observation naturally leads to an important limitation of the present study: this approach only provides local approximations of the dynamics near a local bifurcation point. Accordingly, the scaling laws we derive are expected to hold in a neighbourhood of the bifurcation. As noted by several authors, the correlation between the loss of engineering and ecological resilience can start "far" away from the bifurcation \cite{van2007slow, scheffer2015generic}. This issue ultimately reduces to the meaning of "far" and thus to the choice of appropriate characteristic scales for the system. The validity of the local approximations derived here relies on how far from the bifurcation point the higher-order terms in the Taylor expansion (as in equation~\eqref{eq:fold_DL}) can be neglected to obtain approximate equations (such as equation~\eqref{eq:fold_approx}), and therefore depends on the relative magnitudes of successive partial derivatives of the scalar field.
While this is not a limitation in the context of EWS---which explicitly target systems approaching critical transitions---it raises the question of how resilience metrics behave at larger distances from bifurcation points and whether higher-order approximations could extend the validity of the scalings. 

Finally, our results also contribute to the ongoing debate on the use of polynomial potentials in empirical studies of resilience. Although it has been argued that more justification is needed to fit polynomial potentials to data \cite{dakos2022ecological}, our analysis is a reminder that, locally, any sufficiently smooth nonlinear dynamics can be reduced locally to a polynomial dynamics, which is a consequence of the applicability of Taylor expansions. From this perspective, the use of polynomial potentials is a natural consequence of local expansions theory.

\subsubsection*{Perspective}
We have demonstrated that in the vicinity of local bifurcations, different notions of resilience are strongly coupled, with different patterns per bifurcation type of scalings across resilience metrics. However, as discussed above, we cannot exclude that in the absence of bifurcations, similarly coupled scalings might arise. We see it as a general challenge to combine model determination, the determination of parameters and parameter changes, and assessments of resilience measures, when all this must occur on the basis of time series collected while environments change and bifurcations might occur or not. 


\printbibliography

\appendix

\section{Relationship between curvature and tangent lines in the neighbourhood of a stable equilibrium}\label{an:curvature}
There is a confusion in Box 1 of \cite{dakos2015resilience} between the angle $\varphi$ shown in Figure~\ref{fig:fig3_potential} and the eigenvalue $\lambda_A$ of the stable equilibrium $\eta^*_A$. In this appendix, we clarify the relationship between the local shape of the potential basin in the neighbourhood of $\eta^*_A$ and the value of $\lambda_A$.

The curvature of a function $V$ at a point $x$ is defined as
\begin{equation*}
    K(x) = \frac{V''(x)}{\left(1 + (V'(x))^2\right)^{3/2}},
\end{equation*}
and quantifies the local bending of the curve relative to its tangent at that point.

Accordingly, the curvature of the potential function at the stable equilibrium $\eta^*_A$, characterised by $V'(\eta^*_A)=0$ and $V''(\eta^*_A)>0$, is given by
\begin{equation*}
    K(\eta^*_A) = V''(\eta^*_A) = -\lambda_A,
\end{equation*}
where $\lambda_A < 0$ denotes the eigenvalue associated with the decay of perturbations around $\eta^*_A$.

In the neighbourhood of $\eta^*_A$, for $\delta>0$, the potential admits the local expansion
\begin{equation*}
    V(\eta^*_A \pm \delta)
    = V(\eta^*_A) + \frac{\delta^2}{2} V''(\eta^*_A) + O(\delta^3)
    = V(\eta^*_A) - \frac{\delta^2}{2} \lambda_A + O(\delta^3),
\end{equation*}
showing that the potential is locally parabolic.

\begin{figure}
    \centering
    \includegraphics[width=0.7\linewidth]{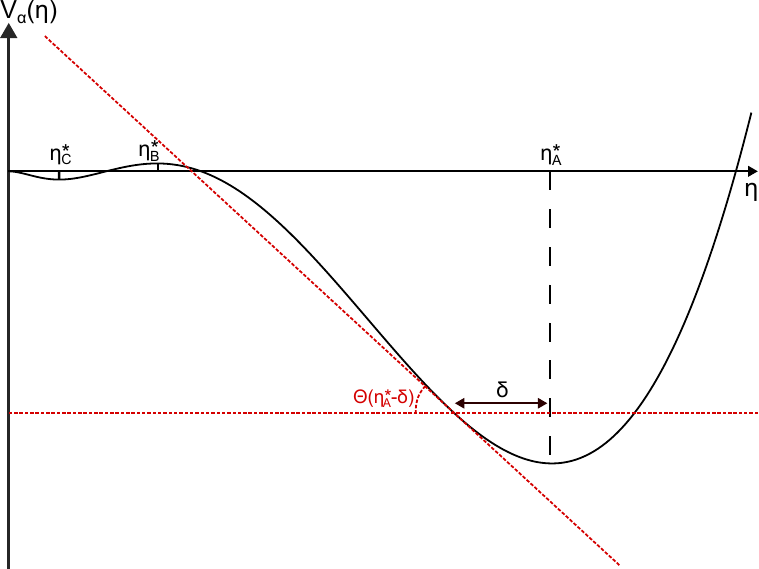}
    \caption{Potential function, as in Figure~\ref{fig:fig3_potential},  with the angle $\Theta$ between the tangent and the horizontal axis indicated.}
    \label{fig:local_angle}
\end{figure}

Let $\Theta(x)$ denote the angle between the tangent to $V$ at $x$ and the horizontal axis, see Figure~\ref{fig:local_angle}. Note that the definition of $\Theta(x)$ could be ambiguous as we do not define its orientation. Therefore, we assume that $\Theta(x) \in [0,\frac{\pi}{2}]$. The following exact relationship holds:
\begin{equation*}
    \tan\left(\Theta(\eta^*_A \pm \delta)\right) = |V'(\eta^*_A \pm \delta)|.
\end{equation*}
Since $V'(\eta^*_A)=0$, we have $\tan\left(\Theta(\eta^*_A \pm \delta)\right) \approx \Theta(\eta^*_A \pm \delta)$ for $0< \delta \ll 1$. Moreover, a Taylor expansion yields
\begin{equation*}
    V'(\eta^*_A \pm \delta) \approx \pm \delta V''(\eta^*_A),
\end{equation*}
and therefore
\begin{equation}\label{eq:curvature}
    \Theta(\eta^*_A \pm \delta) \approx \delta |V''(\eta^*_A) |= \delta |\lambda_A|.
\end{equation}

Therefore, the angle $\Theta$ varies approximately linearly in the vicinity of the stable equilibrium, with a proportionality coefficient given by the amplitude of the dominant eigenvalue. Equation~\eqref{eq:curvature} thus establishes a direct link between the angle of the tangents to the potential curve in the neighbourhood of $\eta^*_A$, the curvature of the potential at the equilibrium, and its dominant eigenvalue.

\section{Explicit derivation of scaling laws in the vicinity of a transcritical bifurcation}\label{an:scalings_transcritical}
Here we explicitly compute how, close to a transcritical bifurcation point,  resilience metrics depend on the distance to the bifurcation point. To this end, we use the approximate dynamics of the process (equation~\eqref{eq:approximate_transcritical_reduced}).

Equation~\eqref{eq:approximate_transcritical_reduced} admits two equilibria for $\eta$, namely $\eta^*+\sigma_\pm(\alpha)$ (see equation~\eqref{eq:expression_sigma} for the expressions of $\sigma_\pm(\alpha)$), so that the distance between them, $ds$, is:
\begin{equation*}
    ds = \lvert \eta^*+ \sigma_+(\alpha) - (\eta^*+\sigma_-(\alpha)) \rvert = \frac{2 \lvert \partial_{\alpha \eta} f( \eta^*, \alpha^*) + \chi'(\alpha^*)\partial_{\eta \eta} f( \eta^*, \alpha^*)\rvert}{\lvert\partial_{\eta \eta} f( \eta^*, \alpha^*)\rvert}  \lvert\alpha-\alpha^*\rvert \propto \lvert\alpha-\alpha^*\rvert
\end{equation*}

We now compute the eigenvalue at the equilibrium $\eta^*+\sigma_+(\alpha)$. Differentiating the right-hand side of equation~\eqref{eq:approximate_transcritical_reduced} with respect to $\eta$ and evaluating it at $\eta = \eta^* + \sigma_+(\alpha)$ yields the eigenvalue of this equilibrium:
\begin{align*}
    \lambda_{\sigma_+(\alpha)} &= \frac{\partial_{\eta \eta}f(\eta^*,\alpha)}{2}\left(\sigma_+(\alpha)-  \sigma_-(\alpha)\right) \\
    &= \frac{\partial_{\eta \eta}f(\eta^*,\alpha)}{2} \frac{2 \lvert \partial_{\alpha \eta} f( \eta^*, \alpha^*) + \chi'(\alpha^*)\partial_{\eta \eta} f( \eta^*, \alpha^*)\rvert}{\partial_{\eta \eta} f( \eta^*, \alpha^*)}  (\alpha-\alpha^*) \\
    &= \lvert \partial_{\alpha \eta} f( \eta^*, \alpha^*) + \chi'(\alpha^*)\partial_{\eta \eta} f( \eta^*, \alpha^*)\rvert (\alpha-\alpha^*).
\end{align*}
Hence $\eta^*+\sigma_+$ is locally stable when $\alpha-\alpha^*<0$ (i.e.\ $\alpha<\alpha^*$) and unstable otherwise, while a similar calculation shows that the reverse holds for $\eta^*+\sigma_-$, which is locally stable when $\alpha>\alpha^*$ and unstable otherwise.
Regarding the dependence on $\lvert\alpha-\alpha^*\rvert$, we therefore obtain:
\begin{equation*}
    \lvert \lambda_{\sigma_\pm(\alpha)}  \rvert \propto  \lvert\alpha-\alpha^* \rvert
\end{equation*}

Finally, we evaluate the potential difference $h$ between the two equilibria:
\begin{align*}
    h &= \vert V_\alpha(\eta^*+\sigma_+(\alpha)) -  V_\alpha(\eta^*+\sigma_-(\alpha)) \rvert  \\
    &= \frac{ \lvert \partial_{\eta \eta}f(\eta^*,\alpha) \rvert }{2} \Big\lvert \frac{\sigma_+^3(\alpha)}{3} - \frac{\sigma_+(\alpha)+ \sigma_-(\alpha)}{2}\sigma_+^2(\alpha) + \sigma_+^2(\alpha) \sigma_-(\alpha) - \frac{\sigma_+^3(\alpha)}{3}  \\
    & \qquad \qquad + \frac{\sigma_+(\alpha) + \sigma_-(\alpha)}{2}\sigma_-^2(\alpha) + \sigma_-^2(\alpha) \sigma_+(\alpha)  \Big \rvert \\
    &= \frac{ \lvert \partial_{\eta \eta}f(\eta^*,\alpha) \rvert }{2} \left\lvert \frac{\sigma_-^3(\alpha) - \sigma_+^3(\alpha)}{6} + \frac{\sigma_+^2(\alpha)\sigma_-(\alpha)- \sigma_+(\alpha)\sigma_-^2(\alpha)}{2}\right\rvert \\
    &\propto \lvert\alpha-\alpha^* \rvert^3
\end{align*}
where this last scaling follows from the linear dependence of $\sigma_\pm(\alpha)$ on $\alpha-\alpha^*$ (see equation~\eqref{eq:expression_sigma}).

\end{document}